\documentstyle[harvard,epsfig]{mn}

\input{epsf}

\def\lsim{\mathrel{\hbox{\rlap{\hbox{\lower4pt\hbox{$\sim$}}}\hbox{$<$}}}}
\def\gsim{\mathrel{\hbox{\rlap{\hbox{\lower4pt\hbox{$\sim$}}}\hbox{$>$}}}}

\def\and   {\rm {et al.} \rm}  

\begin{document}

\title[Distances to Cepheid Open Clusters Via Optical and {\it K}-Band Imaging]
{Distances to Cepheid Open Clusters Via Optical and {\it K}-Band Imaging}

\author[F. Hoyle  et al ]
{
F. Hoyle$^{1,3}$, T. Shanks$^1$, N. R. Tanvir$^{2}$
\\
1 Department of Physics, Science Laboratories, South Road, Durham DH1 3LE
\\
2 Department of Astrophysical Sciences, University of Hertfordshire, College Lane, Hatfield, AL10 9AB\\
3 email fiona.hoyle@durham.ac.uk\\
}

\maketitle

\begin{abstract} We  investigate the reddening and Main Sequence
fitted distances to eleven young, Galactic open clusters that contain Cepheids.
Each cluster contains or is associated with at least one Cepheid variable
star. Reddening to the clusters is estimated using the {\it U-B:B-V}
colours of the OB stars and the distance modulus to the cluster
is estimated via {\it B-V:V} and {\it V-K:V} colour-magnitude diagrams. By
main-sequence fitting  we proceed to calibrate the Cepheid P-L relation and
find M$_V$=-2.81$\times$logP-1.33$\pm$0.32 and M$_K$=-3.44$\times$logP-2.20$\pm$0.29 and a distance modulus to the LMC of 18.55$\pm$0.32 in the {\it V}-band and 18.47$\pm$0.29 in the {\it K}-band giving an overall distance modulus to the
LMC of 18.51$\pm$0.3.

In the case of two important clusters we find that the {\it U-B:B-V} diagram in these clusters is not well fitted by the standard Main Sequence line. In one
case, NGC7790, we find that the F stars show a {\it UV} excess which if
caused by metallicity would imply Fe/H$\sim$-1.5; this is anomalously low
compared to what is expected for young open clusters. In a second case,
NGC6664, the {\it U-B:B-V} diagram shows too red {\it U-B} colours for the F stars which in this case would imply a higher than solar metallicity. If these  effects
{\it are} due to metallicity then it would imply that the Cepheid PL({\it V}) and PL({\it K})
zeropoints depend on metallicity according to $\frac{\delta M}{\delta Fe/H} \sim$0.66 in the sense that lower metallicity Cepheids are intrinsically
fainter. Medium-high resolution spectroscopy for the main-sequence  F stars 
in these two clusters is needed to  determine if metallicity really  is the
 cause or whether some other explanation applies.

Please see http://star-www.dur.ac.uk:80/~fhoyle/papers.html for a version with all the figures correctly inserted.
\end{abstract}


\section{Introduction} Determining the value of Hubble's constant,
H$_{\circ}$, has been a challenge to astronomers since the discovery of the
universal expansion in 1927. It is sometimes argued that we are now at the
fine tuning stage and many measurements give values for H$_{\circ}$ which
lie between the hotly argued values of 50kms$^{-1}$Mpc$^{-1}$ (Sandage) and
100kms$^{-1}$Mpc$^{-1}$ (de Vaucouleurs), e.g. \citeasnoun{tanvir2} calculated
H$_{\circ}$=67$\pm7$kms$^{-1}$Mpc$^{-1}$. However, many of these
measurements are based on secondary indicator methods which in turn are
dependent on the accuracy of  primary indicators of distance such as the
Cepheid Period-Luminosity (P-L) relation. The well-studied LMC P-L relation
is usually calibrated via the distance modulus to the LMC and the previously
accepted value was around 18.50. However, this has been recently challenged
in a paper by \citeasnoun{FC} who determined the distance modulus to the LMC
as 18.70$\pm0.1$. This small difference in the distance modulus causes a
10$\%$ decrease in estimates of the Hubble's Constant. This discrepancy has
further motivated us to check the Galactic zeropoint of the P-L relation. We
do this by checking the values of the distance modulus and reddening of the
11 Galactic clusters that contain Cepheids via zero age main sequence
fitting (ZAMS).

\begin{table*} \begin{tabular}{lcllcccc}   \hline 
Telescope & Date & Cluster & Cepheid & Wavebands & Airmass & Exposure Time(s)  & Photometric \\ 
\hline
JKT 1.0m & 17/9/97 & NGC6823 & SV Vul & {\it UBV} & 1.22 & 2x150, 2x150, 2x150 & No \\ 
JKT 1.0m& 17/9/97 &  & WZ Sgr &  {\it UB} & 1.71 & 3x150, 2x150 & No \\ 
JKT 1.0m& 18/9/97 & M25 & U Sgr &  {\it UBV} & 1.61 & 2x150, 1x150, 1x150 & No  \\ 
JKT 1.0m& 18/9/97 & NGC129 & DL Cas & {\it UBV} & 1.36 & 4x180, 2x180, 1x180 & No \\ 
JKT 1.0m& 20/9/97 & NGC6649 & V367 Sct & {\it UBV} & 1.39 & 4x300, 2x180, 2x120 & No \\
JKT 1.0m& 21/9/97 & NGC6664 & EV Sct &  {\it UBV} & 1.66 & 6x300, 8x300, 10x120 & Yes \\ 
JKT 1.0m& 21/9/97 & Trumpler 35 & RU Sct &  {\it UBV} & 1.22 & 1x180, 1x90,1x90 & Yes \\ 
JKT 1.0m& 21/9/97 & NGC7790 & CEa, CEb, CF Cas & {\it UBV} & 1.29 & 6x300, 8x180, 10x120 &Yes \\ 
\hline 
CTIO 0.9m & 24/9/98 &  & WZ Sgr & {\it UBV} &1.12 & 5x300, 4x150,4x90 & No \\ 
CTIO 0.9m& 27/9/98 & NGC6067 & V340 Nor, QZ Nor& {\it UBV} & 1.36 & 3x300, 1x90, 1x90 & No \\ 
CTIO 0.9m& 27/9/98 & NGC6649 & V367 SCT & {\it UBV} & 1.26 & 5x600, 1x60, 1x60 & No  \\ 
CTIO 0.9m& 28/9/98 & Lynga 6 & TW Nor & {\it UBV} & 1.51 & 4x300, 1x90, 1x90 & Yes \\ 
CTIO 0.9m& 28/9/98 & NGC6067 & V340 Nor, QZ Nor & {\it UBV} & 1.36 & 3x300, 1x90, 1x90 & Yes \\ 
CTIO 0.9m&28/9/98 & vdBergh 1 & CV Mon & {\it UBV} & 1.26 & 1x300, 1x90, 1x90 & Yes \\
CTIO 0.9m& 28/9/98 & NGC6649 & V367 SCT & {\it UBV} & 1.26 & 1x300, 1x60, 1x30 & Yes \\ 
CTIO 0.9m& 28/9/98 & M25 & U Sgr & {\it UBV} & 1.27 & 1x60, 1x5, 1x5&Yes \\
\hline 
UKIRT 3.8m& 16/6/97 & NGC6649 & V367 Sct & {\it K} & 1.40 & 60x2 & Yes \\
UKIRT 3.8m& 17/6/97 & M25 & U Sgr & {\it K} & 1.74 & 60x2 & Yes \\ 
UKIRT 3.8m& 17/6/97 & Trumpler 35 &  RU Sct & {\it K} & 1.64 & 60x2 & Yes \\ 
UKIRT 3.8m& 19/6/97 & NGC6664 & EV Sct & {\it K} & 1.14 & 60x2 &Yes \\ 
UKIRT 3.8m& 19/6/97 & NGC6823 & SV Vul & {\it K} & 1.03 & 60x2 &Yes \\ \hline 
Calar Alto 3.5m& 18/8/97 & NGC129 & DL Cas & {\it K$_{\rm short}$} & 1.25 & 10x1.5 &Yes \\ \hline 
WHT 4.2m& 1/9/96 & NGC7790 & CF, CEa, CEb Cas & {\it K$_{\rm short}$} & 1.205 & 50x1 & Yes \\ \hline 
\end{tabular} 
\caption{Details of the observations. The airmass is the average airmass of the exposure and the exposure time is given in seconds for each of the wavebands in column 5.} 
\label{tab:data} 
\end{table*}

Previous work on measuring the reddening and distance to young open clusters
which contain Cepheids via ZAMS fitting has been done using photoelectric
and photographic measurements in optical wavebands. It is time consuming to
observe a large number of stars using photoelectric observations as each
star has to be observed individually. Photographic data can give relatively
inaccurate magnitudes and colours. However, CCD's now make it possible to
observe a large number of stars in many different wavebands quickly and
accurately. Although CCD's have already been used for open cluster studies
e.g. \citeasnoun{walkly6}, \citeasnoun{walk6067}, \citeasnoun{romeo}, these
have mainly been carried out in {\it BVRI}. Recently CCD's with improved {\it U}-band
sensitivity have become available and {\it U}-band CCD data is included in this
study. Infra-red imaging detectors are also now available and although some
of the detectors used here do not cover as wide an area as optical CCD's, observing the full extent of an open cluster with a mosaic of pointings is a practical proposition.

Until fairly recently, good quality infra-red measurements of the Cepheids
themselves were not available and the Cepheid P-L relation has been
primarily calibrated in the {\it V}-band. Laney and Stobie (1993,1994) present
infra-red along with {\it V}-band magnitudes for a large number of Southern
Hemisphere Galactic Cepheids. Using data in the literature to obtain values
for the distance modulus and reddening to the clusters they calibrated the
Cepheid P-L relation in the {\it V} and {\it K}-band. Any errors in the determination
of the distance modulus and the reddening in the previous work would cause
an error in the PL relation as determined by Laney and Stobie.

The layout of this paper is as follows. In section \ref{sec:data} we present
the observational data and we test the accuracy of the photometry and
calibration of the data. In section \ref{sec:method} we describe how the
reddenings and distances to the open clusters are obtained and in section
\ref{sec:clus} we discuss each cluster individually. In section \ref{sec:PL}
we use these values with the magnitudes of the Cepheids to calibrate the
Cepheid Period-Luminosity relation, In section \ref{sec:interp} we discuss the implications of the results, particularly for the clusters whose {\it U-B:B-V} diagrams do not appear to follow the canonical locus. In section \ref{sec:conc} we draw conclusions.

\section{Data} \label{sec:data}

\subsection{Observations} \label{sec:obs}

The observations of the Galactic Open clusters were taken during five
observing runs on the JKT, UKIRT, at CTIO, at Calar Alto and on the WHT over
a two year period. The spread in declination of the clusters and the
multi-wavelength nature of the study meant that many different telescopes
were required.

\subsubsection{JKT}

Optical imaging of eight open clusters was obtained during an observing run
from the 16/9/1997 to the 22/9/1997. The observations were carried out using
the 1024$\times$1024 Tektronix CCD with pixel scale of 0.33 arcsec pixel$^{-1}$. Typical
seeing was around 1.3$^{\prime\prime}$. Short exposures of 5s in {\it V}, 10s in {\it B}
and 20s in {\it U} were observed for calibration purposes but the main imaging
observations were typically 6x120s in the {\it V}-band, 6x180s in the {\it B}-band
and 6x300s in the {\it U}-band. Due to the Southerly declination of some of the objects, they
had to be observed at high air mass. However these observations were
normally used to obtain relative photometry and calibration frames were
observed at as low an airmass as possible or during a later observing
run at CTIO. The pointings are given in Table \ref{tab:point}. For calibration purposes, standard stars from \citeasnoun{landolt} were used. On the one
fully photometric night (21/9/97) six Landolt fields were observed at
regular intervals throughout the night, most of these fields containing
several standard stars.

\begin{table}
\begin{tabular}{ll} 
Cluster & Pointing \\ \hline
NGC6649 & Star 19 in \protect\citeasnoun{MVDB} \\
M25 & Star 95 in \protect\citeasnoun{sandm25} \\
NGC6664 & Star 5 in \protect\citeasnoun{arpEV} \\
WZ Sgr & WZ Sgr \\
Lynga 6 & TW Nor \\
NGC6067 & Star 136 in \protect\citeasnoun{thack} \\
vdBergh 1 & CV Mon \\
TR35 & 5$^{\prime\prime}$ south of TR35 \\
NGC6823 & Star j in \protect\citeasnoun{guetter} \\
NGC129 & Star 113 in \protect\citeasnoun{arp129} \\ 
NGC7790 & Star E in \protect\citeasnoun{romeo} \\ \hline
\end{tabular}
\caption{Approximate pointings for the clusters in the study in all wavebands.}
\label{tab:point}
\end{table}

\subsubsection{CTIO} The observations were made  using the CTIO 0.9-m during
an observing run from the 24/9/98 to 29/9/98. These observations were
carried out using the 2048x2048 Tek\#3 CCD with pixel scale 0.384 arcsec
pixel$^{-1}$. The Tek\#3 CCD has low readout noise (4 electrons) and good quantum efficiency in the {\it U}-band. The average seeing during the observations was
around 1.4$^{\prime\prime}$. Approximate pointings are again given in Table \ref{tab:point}. Standard E-region fields from
\citeasnoun{graham} and standard stars from \citeasnoun{landolt} were
observed throughout the night. The only photometric night was  the 28/9/1998
and on this night six E-region standard fields and one Landolt standard
field containing four standard stars were observed throughout the night.
Three new clusters were observed at CTIO and further observations of
clusters observed at the JKT were made in cases where the clusters had been observed in non-photometric conditions only.

\subsubsection{UKIRT}

The infra-red data was mostly observed at UKIRT during the four nights
16-19/6/97 using the IRCAM3  near-IR imaging camera with a 256x256 detector.
The pixel scale used was 0.286 arcsec pixel$^{-1}$ giving a field of view of
73$^{\prime\prime}$. To cover a sufficient area of each open cluster we therefore had to
mosaic images. Generally a mosaic of 9x7 images was observed. Each image was
overlapped by half in both the x and y direction so the final image had an
approximate area of 6$^{\prime}\times$ 5$^{\prime}$. Observations were generally taken using
the ND-STARE mode, where the array is reset and read immediately and then
read again after the exposure which reduces the readout noise  to 35e$^{-}$.
The exposures were 60$\times$2 seconds and the centre of the mosaic is approximately in the same position as the corresponding optical frame. Standards from the UKIRT faint
standards list  were observed throughout the nights. Around 10 standards
were observed on the three photometric nights, some of which were observed
early on in the night, half way through and at the end of the night. The
seeing throughout the run was typically 0.6$^{\prime\prime}$. Observations
were also taken in the {\it J} and {\it H}-band.

\subsubsection{Calar Alto}

NGC129 lies further north than the declination limit of UKIRT so infra red observations were instead
taken at Calar Alto during another observing run. Observations were done
using the Rockwell 1k$\times$1k Hawaii detector with pixel scale 0.396 arcsec pixel$^{-1}$.
This gives a 6.$^{\prime}$6 field of view so there was no need for the mosaicing
technique used at UKIRT. The exposure time was 10$\times$1.5s, the seeing
was better than 1$^{\prime\prime}$ and the exposure was centred on star 113 in \citeasnoun{arp129}. Observations were made in the {\it K$_{\rm short}$}-band. UKIRT faint standards and standards of Hunt et al (1998) were observed for calibration.

\subsubsection{WHT} 

NGC7790 also lies further north than the UKIRT declination limit and so infra red observations were made on the 4.2m WHT during another observing run. The observations were made on the 1/9/1996. The WHIRCAM 256$\times$256 detector which was
situated at the Nasmyth focus and behind the MARTINI instrument was used for the observations, without MARTINI tip-tilt in operation.
The WHIRCAM detector was the IRCAM detector previously used at
UKIRT. The pixel size was 0.25 arcsec pixel$^{-1}$ and the field-of-view was therefore 64$^{\prime\prime}$, centred on star E in \citeasnoun{romeo}. The {\it K$_{\rm short}$} filter was used and UKIRT faint standards were
observed for calibration.

\subsubsection*{} All the observations are summarized in Table
\ref{tab:data}. The date of each observation, the wavebands observed for
each cluster and the airmass are given in columns 2, 5 and 6 respectively.
Column 7 gives the exposure time of the frames used for imaging. For some of
the clusters a calibration frame was observed at CTIO and the exposure time
of these clusters are also given in column 7. Column 8 indicates where a
cluster was observed on a photometric night and hence where an independent
zero point was obtained.

\begin{table*}
\begin{tabular}{lccccl} 
Cluster & {\it U}$_{\rm{obs}}$ - {\it U}$_{\rm{previous}}$ & {\it B}$_{\rm{obs}}$ - {\it B}$_{\rm{previous}}$  & {\it V}$_{\rm{obs}}$ - {\it V}$_{\rm{previous}}$ & N(stars) &Previous Work \\ \hline
NGC6649 & 0.02$\pm$0.025 & 0.01$\pm$0.01 & 0.00$\pm$0.012 & 20 & \protect\citeasnoun{MVDB} \\
M25 & -0.01$\pm$0.03 & -0.03$\pm$0.02 & -0.02$\pm$0.02 & 21 & \protect\citeasnoun{sandm25} \\
NGC6664 & -0.01$\pm$0.012 & -0.02$\pm$0.01 & -0.01$\pm$0.015 & 15 & \protect\citeasnoun{arpEV} \\
Lynga 6 & 0.02$\pm$0.02 & 0.02$\pm$0.02 & -0.01$\pm$0.015 & 17 & \protect\citeasnoun{moff} \\
NGC6067 & 0.07$\pm$0.08 & 0.04$\pm$0.03 & -0.06$\pm$0.04 & 6* & \protect\citeasnoun{thack} \\
vdBergh 1 & 0.02$\pm$0.015 & 0.00$\pm$0.02 & -0.005$\pm$0.008 & 24 & \protect\citeasnoun{arpCV} \\
Trumpler 35 & -0.04$\pm$0.02 & -0.01$\pm$0.015 & 0.01$\pm$0.015 & 15 & \protect\citeasnoun{hoag} \\
NGC7790 & -0.009$\pm$0.01 & -0.015$\pm$0.008 & 0.01$\pm$0.007 & 22 & \protect\citeasnoun{ped} \\ \hline
WZ Sgr & 0.00$\pm$0.01 & 0.00$\pm$0.011 & 0.00$\pm$0.007 & 12 & \protect\citeasnoun{turnWZ}\\
NGC6823 & 0.00$\pm$0.01 & 0.00$\pm$0.005 & 0.00$\pm$0.004 & 13 & \protect\citeasnoun{guetter} \\
NGC129 & 0.00$\pm$0.03  & -0.005$\pm$0.01 & -0.008$\pm$0.01 & 13 &  \protect\citeasnoun{turn129} \\ \hline
\end{tabular}
\caption{Comparison of the photometry of this study with previous photoelectric data. Note there are 6 stars from \protect\citeasnoun{thack} in common with the {\it B} and {\it V}-band data but only 2 in common with the {\it U}-band data. The last three clusters were observed on non-photometric nights only so the previous work was relied upon for calibration. There is a small offset between the zero point of NGC129 and \protect\citeasnoun{turn129} as only the brightest stars were used for calibration.}
\label{tab:resids}
\end{table*}

\subsection{Data Reduction} 
\subsubsection{JKT} 
\label{sec:jktred}

Removal of the bias introduced into the data and trimming of the frames to
remove the overscan region was done on all the frames using the IRAF task
CCDPROC. At least eight {\it U}-band sky flats and six {\it B} and {\it V}-band sky
flats were observed on each of the nights so a separate flat field was
created for every night using a combination of dust and dawn sky flats. This
was created within FLATCOMBINE using a median combining algorithm and a
3$\sigma$ clipping to remove any cosmic rays. The residual gradient in the
flat fields is around 1\%. The task CCDPROC then applies the flat fields to
all the images. The same flat fields were used in the reduction of the
standard star frames.

Many images of the same cluster were observed. These were all combined
together by aligning the images with linear shifts using the task IMSHIFT.
Generally these shifts were small (a few pixels either way) as the
observation were done one after each other and in some cases no shifts were
required. The images were combined using IMCOMBINE and were averaged
together using a 5$\sigma$ clipping.

\subsubsection{CTIO} 
\label{sec:ctiored}

The data was obtained at CTIO using the four amplifier readout mode. A
package called QUADPROC within IRAF corrects for the different bias levels
in the four quadrants of the CCD. The frames were also trimmed using
QUADPROC. In a previous observing run, \citeasnoun{croomphot} had found that
there was a residual gradient of 5\% in the dome flat fields so sky flats
were used. At least three sky flats were observed on each night in the {\it B} and
{\it V}-band and at least 5 sky flats were observed on each night in the {\it U}-band.
The resulting flat fields were flat to better then 1\%. The equivalent
version of FLATCOMBINE in the QUAD package was used to median combine the
flat fields using a 3$\sigma$ clipping. The E-region standards and
observations of Landolt standards were again reduced in the same manner.

Multiple images of the same cluster were again combined using IMCOMBINE with
the same settings as for the JKT data and where any offset shifts appeared
between the data frames they were again corrected for using IMSHIFT.

\subsubsection{UKIRT} 
\label{sec:ukirtred}

The UKIRT data was reduced using a program called STRED within the package
IRCAMDR. This is a fairly automated routine which reads in the data frames,
subtracts of the dark count and creates a flat field frame by median
filtering the image frames. Then the program flat fields the dark subtracted
object images, corrects for any bad pixels and finally creates a mosaic.
All the data frames were median combined to create a flat field for each night. There was no evidence of a large scale gradient greater than about 1\% in the flat fields.

To create the final image, all the individual frames have to be mosaiced
together. STRED reads in the offset from the data header, however, these
offsets were not accurate enough. By creating a separate offsets
file the mosaicing could be done more accurately. To create the offsets file, one of the corner frames was fixed and the offset required for the neighbour frame were found by eye. This was built up over the whole frame, however once one offset had been determined, all the other offsets in the x and y direction from frame to frame were the same. The offsets for the standard stars were more accurate and
could be used to create the mosaic.

\subsubsection{Calar Alto} \label{sec:calarred}

Basic IRAF routines such as IMCOMBINE and IMARITH were used to reduce the
Calar Alto data. A flat field was created by median combining all the data
frames. A more detailed description of the data reduction can be found in
\citeasnoun{henry}. We found that the best results were obtained by first
subtracting a sky frame from each image, as follows: A sky frame for each
individual image was created using IMCOMBINE with a $5\sigma$ clipping to
median filter four data frames that were local in time to the image frame to
create a sky frame for each data frame. The sky frame was then subtracted
off the image before IMCOMBINE was used again to combine the data frames,
forming one final image frame.

\subsubsection{WHT} \label{sec:whtred}

The data from the WHT was reduced for us as part of another project.
Dome flats were used to flat field the data and this was divided into the science frame using IMARITH. Sky subtraction was also required. A sky frame was created by combining dedicated sky frames observed locally in time to the science frame and this was then subtracted from the science frame also using IMARITH.

\begin{figure*} \begin{tabular}{ccc} 
{\epsfxsize=5.6truecm\epsfysize=6truecm\epsfbox[60 170 550 620]{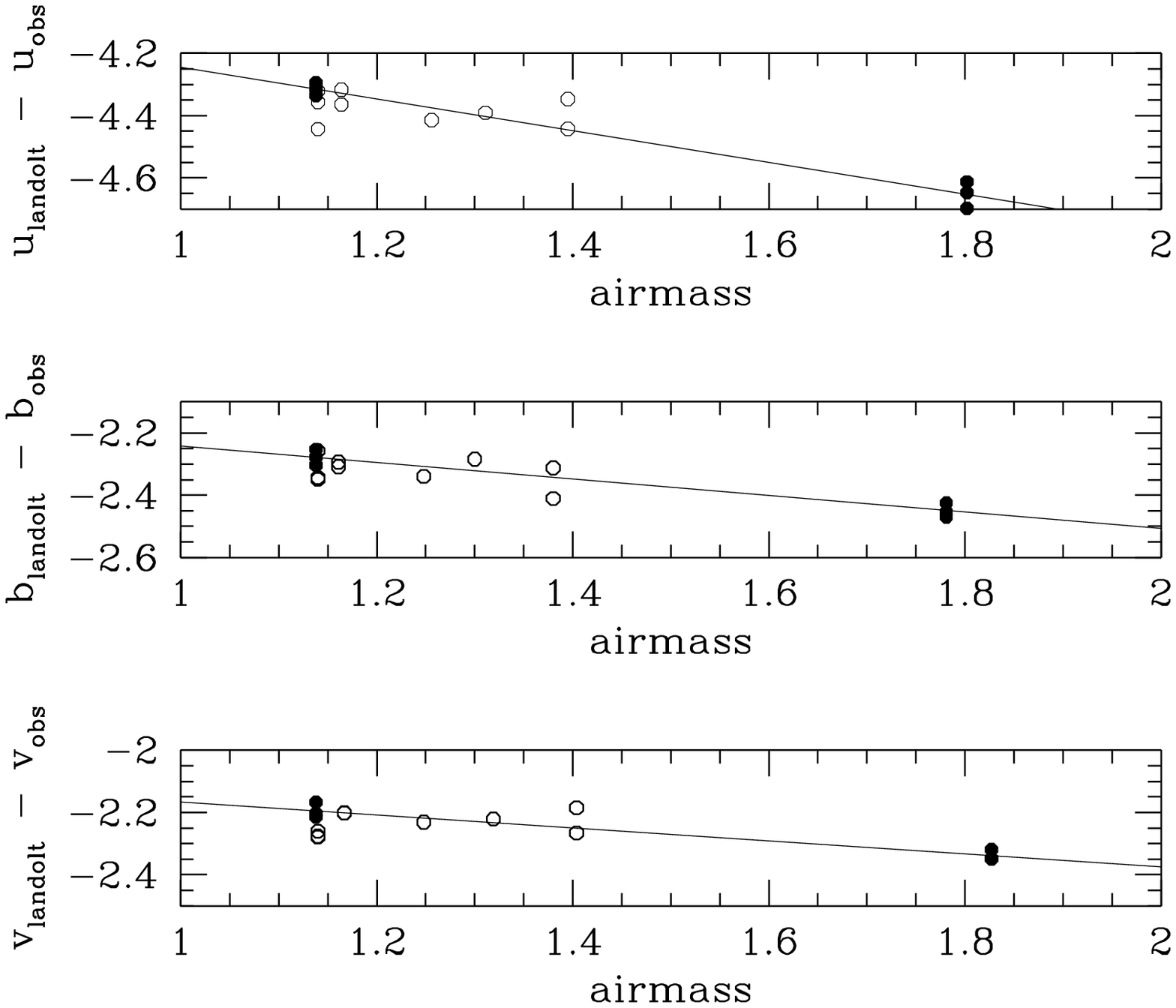}} &
{\epsfxsize=5.6truecm\epsfysize=6truecm\epsfbox[60 170 550 620]{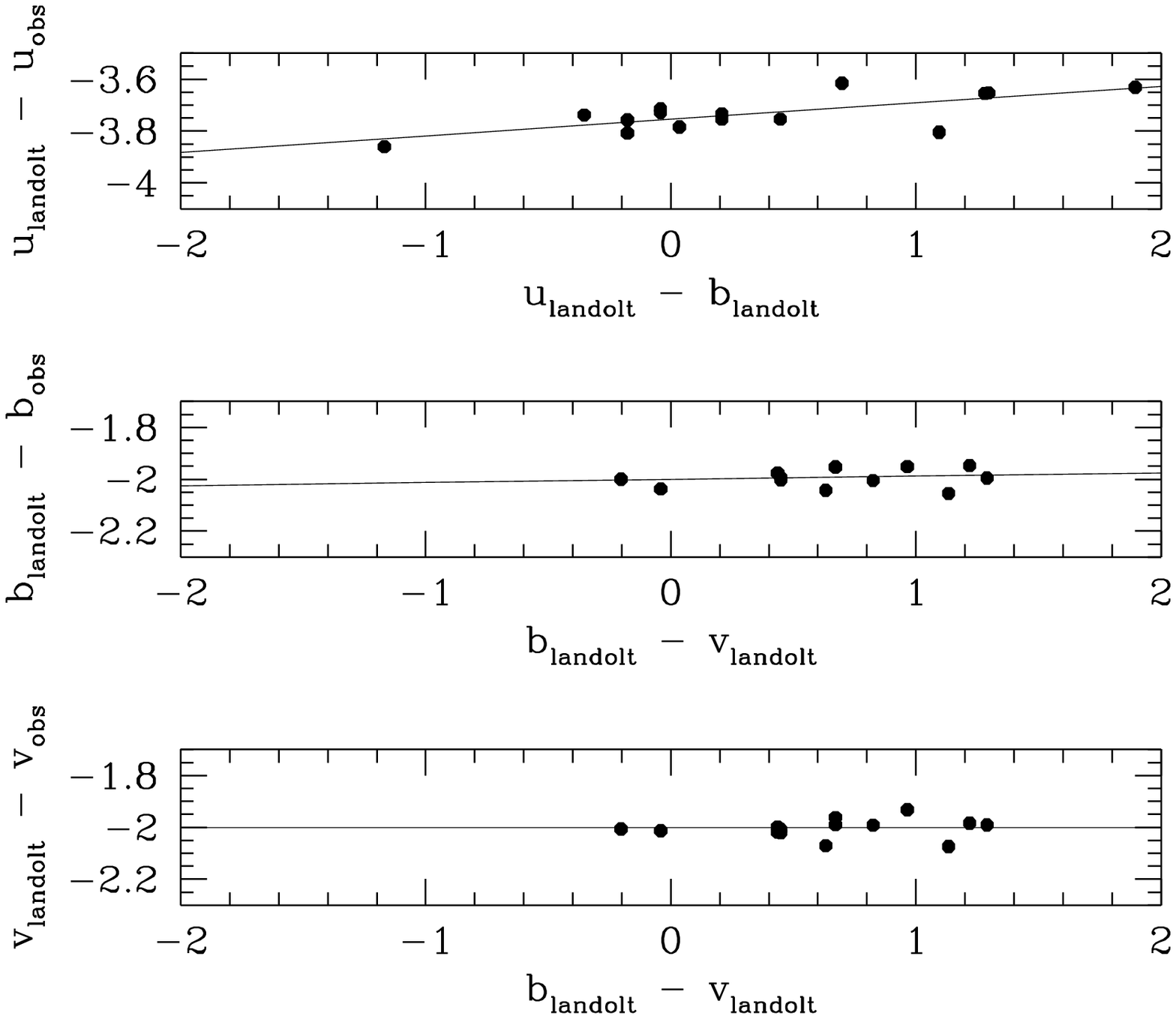}} & 
{\epsfxsize=5.6truecm\epsfysize=6truecm\epsfbox[60 170 550 620]{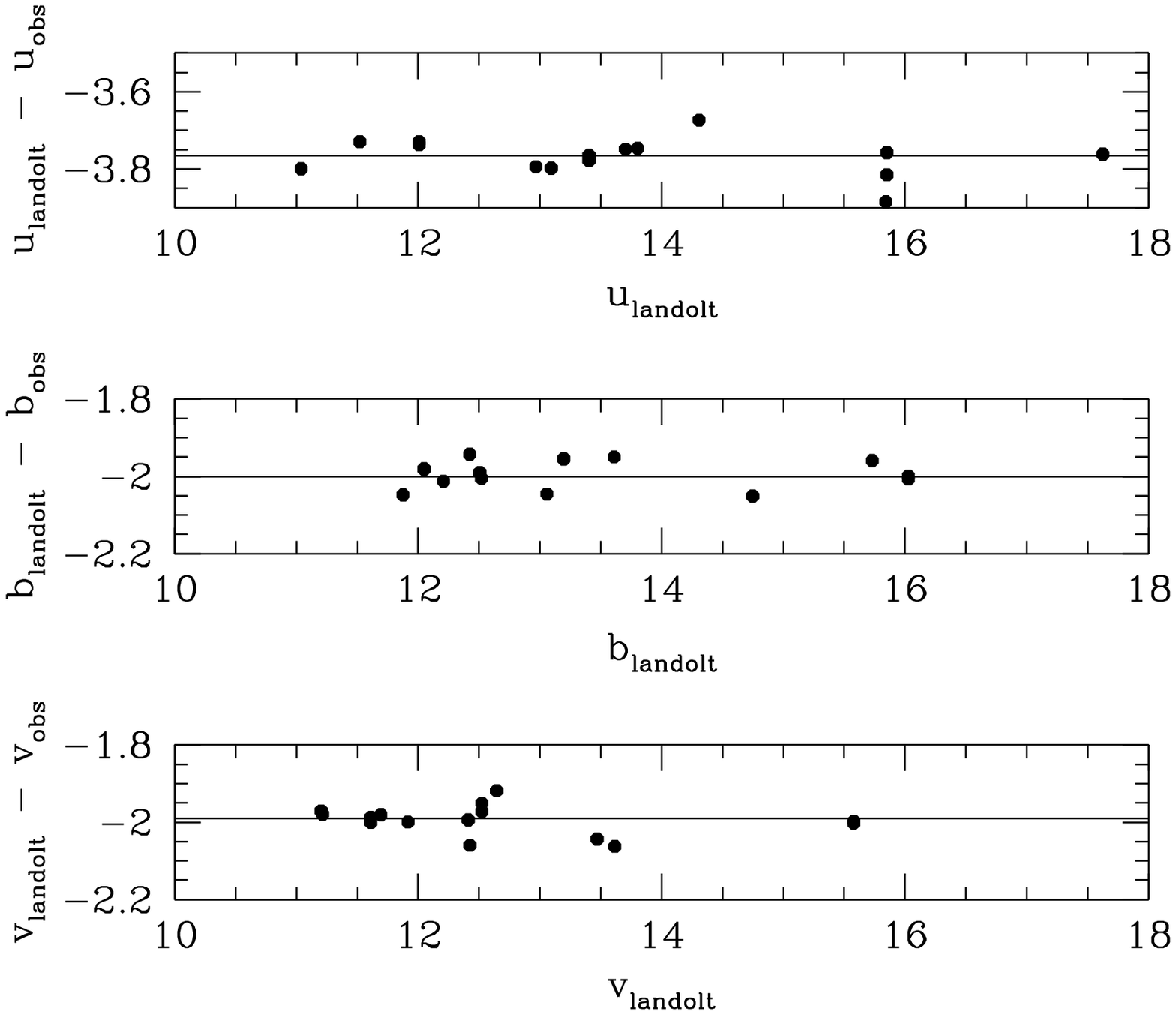}} \\ 
\end{tabular}
\caption{The airmass coefficient, colour equation and zero point, with the
colour equation and airmass correction applied, for the {\it U}, {\it B} and {\it V} wavebands for the data from JKT. The open circles in the airmass plot show the standard stars observed at one airmass only.} 
\label{fig:zerojkt} 
\end{figure*}

\subsection{Image Alignment}

In order to be able to produce colour-magnitude diagrams, the magnitude of
each star in all the different wavebands is required. To do this, all the
frames need to be aligned. Aligning the optical frames was easy as there
were only linear shifts between each waveband. Rather than altering the
data, the alignment was just done by applying small corrections to the x and
y positions of the stars. Aligning the optical data with the {\it K}-band data
was more complicated though as there were shifts, shears and rotations
between the frames. We used the IRAF routine GEOMAP to calculate the best
spatial transformation function between any two images thus allowing the
optical and {\it K}-band data to be aligned. The mapping was only used to
transform coordinates and we did not perform photometry on the resampled
images.

\subsection{Photometric Calibration} \label{sec:cal}

\begin{figure*} 
\begin{tabular}{ccc} 
{\epsfxsize5.6truecm \epsfysize=6truecm
\epsfbox[60 170 550 620]{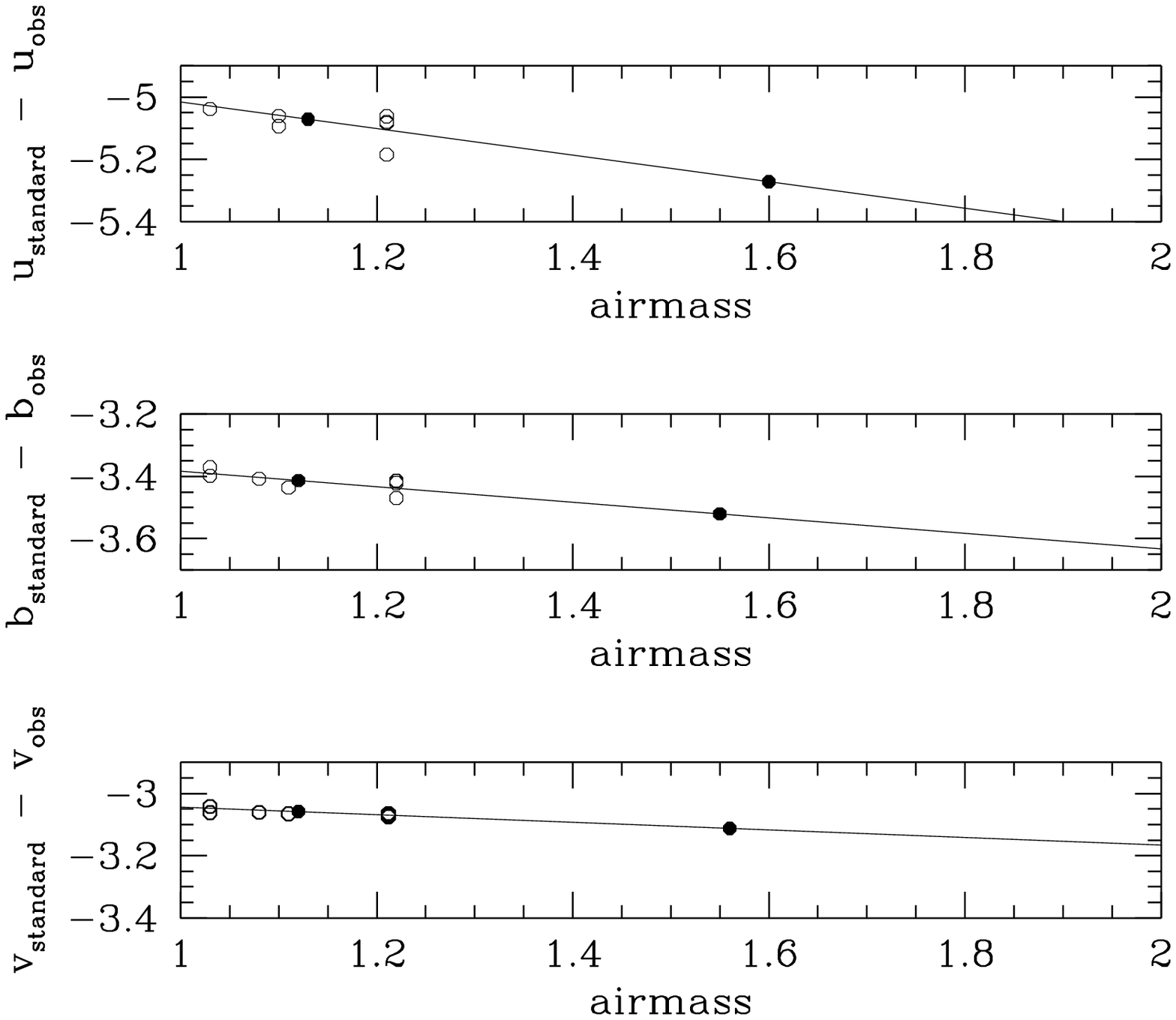}} &
{\epsfxsize=5.6truecm \epsfysize=6truecm \epsfbox[60 170 550
620]{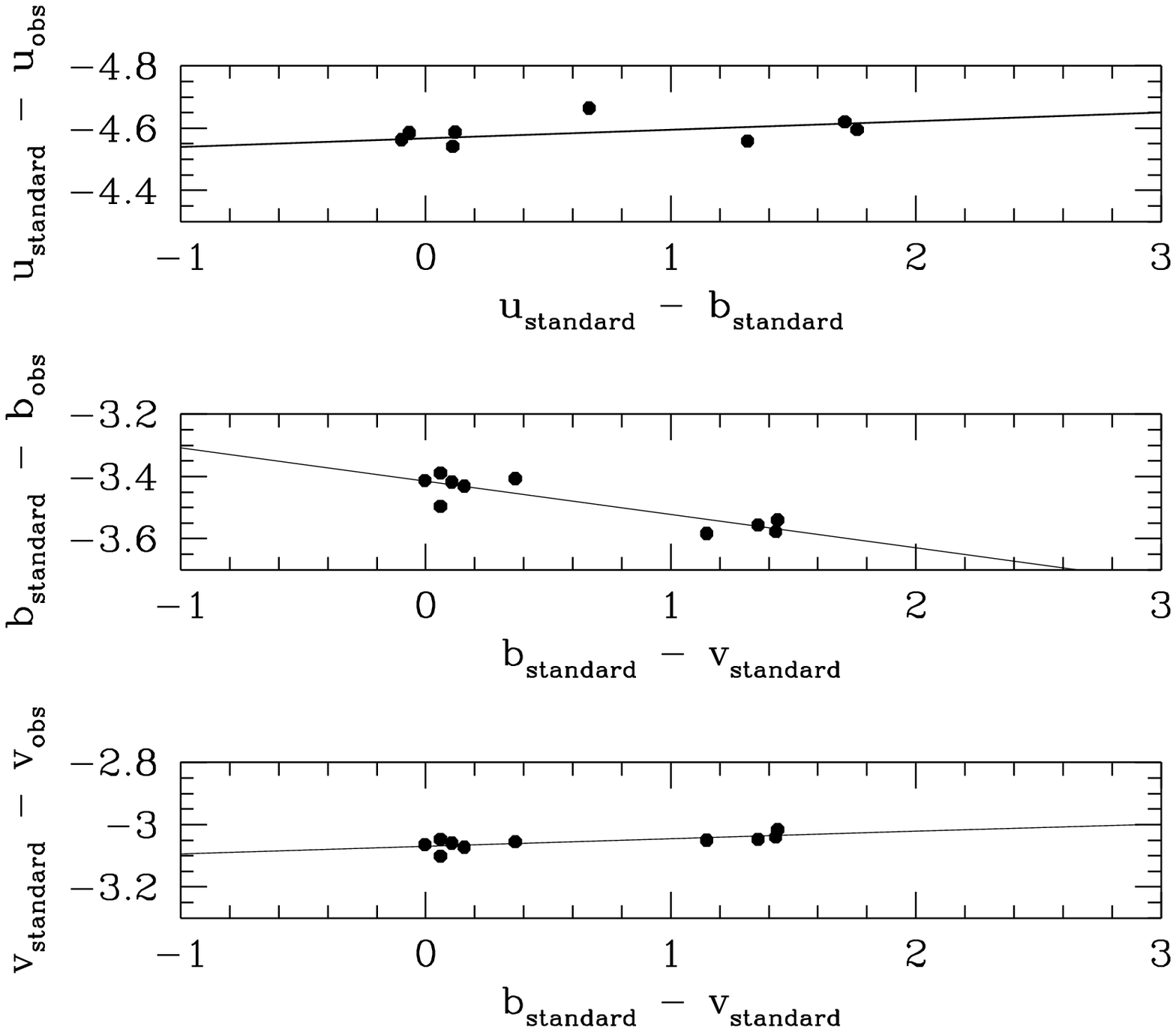}} & 
{\epsfxsize=5.6truecm\epsfysize=6truecm \epsfbox[60 170 550
620]{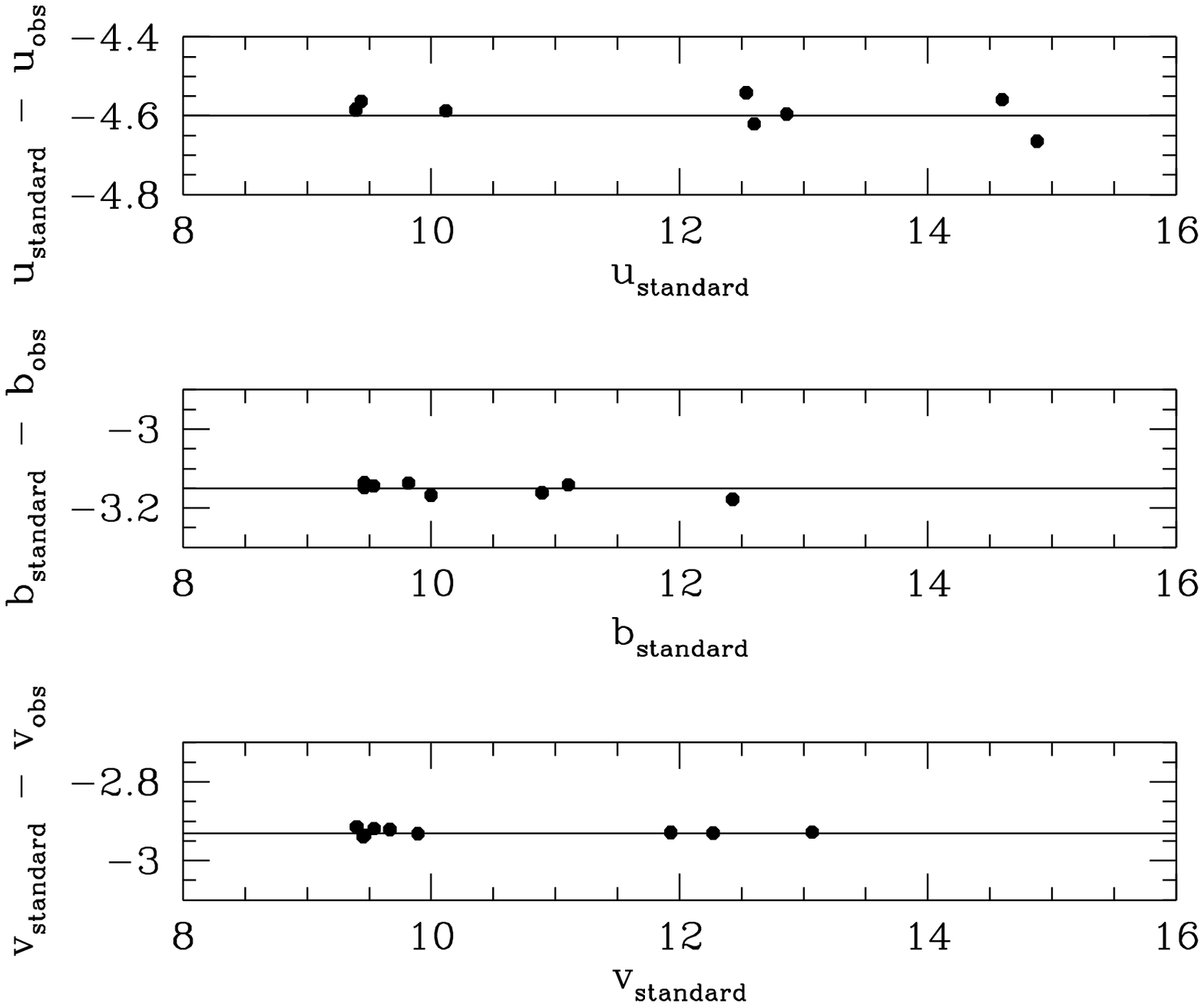}} \\ 
\end{tabular}
\caption{The airmass coefficient, colour equation and zero point, with the
colour equation and airmass correction applied, for the {\it U}, {\it B} and {\it V}
wavebands for the data from CTIO. The open circles in the airmass plot show the standard stars observed at one airmass only.} 
\label{fig:zeroctio} 
\end{figure*}

\subsubsection{JKT} 
\label{sec:jktcal}

Observations of the standard stars of \citeasnoun{landolt} were taken at
regular intervals on each of the six nights at the JKT. The CCD frames of
the standard stars were reduced in the same manner as the data frames with
the same flat fields etc as discussed in section \ref{sec:jktred}. The
aperture size used to measure the magnitude of the standard stars was
15$^{\prime\prime}$, large enough to  determine accurately the total
magnitude of the star but not so large that sky subtraction errors dominate
the magnitude measurement.

Only the sixth night (21/9/97) was fully photometric and the zero point,
airmass coefficient and colour equation are shown in Figure
\ref{fig:zerojkt} and given below with the rms scatter.

\begin{displaymath}
U_{\rm{jkt}} = U_{\rm{ldt}} + 3.76 + 0.49sec(z) - 0.063(U_{\rm{ldt}}-B_{\rm{ldt}}) \pm0.038
\end{displaymath}
\begin{displaymath}
B_{\rm{jkt}} = B_{\rm{ldt}} + 2.00 + 0.27sec(z) - 0.013(B_{\rm{ldt}}-V_{\rm{ldt}}) \pm0.033
\end{displaymath}   
\begin{displaymath}
V_{\rm{jkt}} = V_{\rm{ldt}} + 1.99 + 0.19sec(z)  \pm0.026
\end{displaymath}
where the subscript ldt stands for the Landolt
standard star magnitude and the subscript jkt stands for the instrumental
magnitude, z represents the zenith distance. The errors on the airmass are $\pm$0.0025, 0.0012 and 0.0008 in the {\it U, B} and {\it V} and the errors on the colour equation are $\pm$0.0007 and 0.0005 in the {\it U} and {\it B}-bands respectively. There are no Landolt magnitudes available for the data frames so the colour term has to be translated in instrumental magnitudes. The colour term is negligible in the {\it V}-band calibration so the instrumental {\it V}-band magnitudes come directly from the above equations. The {\it B}$_{\rm ldt}$-{\it V}$_{\rm ldt}$ and {\it U}$_{\rm ldt}$-{\it B}$_{\rm ldt}$ colours are given by
\begin{displaymath}
(B_{\rm{ldt}}-V_{\rm{ldt}}) = 1.013((B_{\rm{jkt}}-V_{\rm{jkt}}) - 0.01 - 0.08sec(z))
\end{displaymath}
\begin{displaymath}
(U_{\rm{ldt}}-B_{\rm{ldt}}) = 1.067((U_{\rm{jkt}}-B_{\rm{jkt}}) - 1.76 - 0.22sec(z))
\end{displaymath}
we assume that the contribution from the colour term in the calibration of the {\it B}-band is negligible when determining the {\it U}$_{\rm ldt}$-{\it B}$_{\rm ldt}$ colour. These exact colour terms are used to correct the instrumental magnitudes in order to make the colour-colour and colour-magnitude diagrams in Figures \ref{fig:ubbv}, \ref{fig:bv} and \ref{fig:vk}.

The clusters NGC7790, NGC6664 and Trumpler 35 were observed on the one
photometric night. Short exposure observations of NGC6649 and M25 were made
at CTIO in order to obtain an independent zero point for these frames.
Around 20 bright (brighter than {\it V}$\sim$15), fairly uncrowded, unsaturated stars were
taken as standard stars to identify the relation between the zero point from
the JKT data and from CTIO data to an accuracy of 0.01 mags. The agreement between the zero point found via this method and previous, photoelectric calibrations is good (see Table \ref{tab:resids}) with only small offsets in each case. The remaining three clusters were observed in non-photometric conditions only so previous work had to be relied upon
for the calibration. For NGC6823 the photoelectric observations from Table 1
of \citeasnoun{guetter} were used. Some of these stars were saturated on the
CCD frame and the area of overlap between the two images was not identical but
thirteen stars were suitable for calibration purposes. \citeasnoun{guetter} compares his photoelectric data with that of previous work and finds good agreement. Two sources of photoelectric data are available for the cluster NGC129, \cite{arp129} and \cite{turn129}. There are 9 stars in common with the Arp photometry and 13 stars in common with the Turner photometry. For these samples of stars, we find that the {\it U}-band zero point obtained from Arp is 0.04$\pm$0.03 mags brighter than that of Turner. In the {\it B}-band the difference is 0.02$\pm0.01$ mags in the sense that Arp is brighter than Turner and there is 0.01$\pm0.01$ mag difference in the same sense in the {\it V}-band. These differences are mainly caused by the stars in the sample with {\it V} fainter than 14 mag. We therefore use the average of the brightest two stars from Arp and Turner to calibrate NGC129. The zero point from this method agrees very well with the zero point obtained using Turner's photometry, which is shown in Figure \ref{fig:other}. Finally the photoelectric work of \citeasnoun{turnWZ} was used to calibrate the cluster containing the Cepheid WZ Sgr. 

The magnitudes of the stars on the data frames on all nights were measured using a 5$^{\prime\prime}$ aperture. The standard star magnitudes were measured using a 15$^{\prime\prime}$ aperture so for the photometric night (21/9/97) an aperture correction had to be applied to the data. The correction in the {\it U}, {\it B} and
{\it V}-bands are -0.165$\pm$0.02, -0.11$\pm$0.025 and -0.11$\pm$0.024 magnitudes respectively as determined by comparing the magnitudes of the standard stars with a 5$^{\prime\prime}$ and 15$^{\prime\prime}$ aperture. No aperture correction was required for those cases where the calibration was tied to a photometric sequence in the cluster itself. For the clusters NGC6649 and M25 the CTIO aperture correction was required.

\subsubsection{CTIO} 
\label{sec:ctiocal}

E-region standards from \citeasnoun{graham} and Landolt standards were
observed for the photometric calibration of the optical CTIO data.
\citeasnoun{menz} compared the zero points and colour differences found from
using the two different standard star studies and found the offsets between
the two to be small, 0.004$\pm$0.0095 offset in the sense E regions -
Landolt. and similar sized offsets in the {\it U-B} and {\it B-V} colours. Any
offsets are within the quoted error. As before, these standard frames were
reduced in the same manner as the data frames and a 15$^{\prime\prime}$
aperture was used to determine the magnitude. Again, only one night was
photometric and this was the last night (28/9/98). The zero points, airmass
coefficients and colour equations for each waveband, {\it U}, {\it B} and {\it V}, are
shown in Figure \ref{fig:zeroctio} and given below again with the rms scatter.

\begin{displaymath} 
U_{\rm{ctio}} = U_{\rm{std}} + 4.61 + 0.47sec(z) -
0.036(U_{{\rm std}}-B_{{\rm std}}) \pm0.035 
\end{displaymath}
\begin{displaymath} 
B_{\rm{ctio}} = B_{\rm{std}} + 3.15 + 0.21sec(z) +
0.099(B_{\rm{std}}-V_{\rm{std}}) \pm0.015 
\end{displaymath}
\begin{displaymath} V_{\rm{ctio}} = V_{\rm{std}} + 2.93 + 0.12sec(z) -
0.018(B_{\rm{std}}-V_{\rm{std}}) \pm0.007 
\end{displaymath} 
where the subscript std stands for the E-field or Landolt standard
magnitude and ctio stands for the instrumental magnitude. The errors on the airmass are $\pm$0.003, 0.0011 and 0.0009 in the {\it U, B} and {\it V} and the errors on the colour equation are $\pm$0.0006, 0.0007 and 0.001 in the {\it U}, {\it B} and {\it V}-bands respectively. There is
generally good agreement between the values for the airmass coefficients
and colour equations found in this work and in \citeasnoun{croomphot}. Again, a 5$^{\prime\prime}$ aperture was used for the data frames so an aperture correction for the photometric night was required. The aperture corrections, in the {\it U}, {\it B} and {\it V}-bands, are -0.17$\pm$0.03, -0.17$\pm$0.02 and -0.13$\pm$0.02 magnitudes respectively. Again, for the data frames the colour terms have to be found in terms of CCD magnitudes rather than standard magnitudes. The colour term in the {\it B}-band is in this case non-negligible so {\it V}-band magnitudes have to be used in the {\it U}-band calibration. The standard colours are given by
\begin{displaymath}
(B_{\rm{std}}-V_{\rm{std}}) = 1.088((B_{\rm{ctio}}-V_{\rm{ctio}}) - 0.22 - 0.09sec(z))
\end{displaymath} 
\begin{displaymath}
\begin{array}{r}\!\!\!(U_{\rm{std}}-B_{\rm{std}}) = 1.037((U_{\rm{ctio}}-B_{\rm{ctio}}) - 1.46 - 0.26sec(z) \\ + 0.099(1.088((B_{\rm{ctio}}-V_{\rm{ctio}}) - 0.22 - 0.09sec(z)))) \end{array}
\end{displaymath} 
again these colours are used in all to calculate the instrumental magnitudes which are used in the colour-colour and colour-magnitude diagrams in Figures \ref{fig:ubbv}, \ref{fig:bv} and \ref{fig:vk}.

This night provided independent zero points for the clusters NGC6067, Lynga
6 and vdBergh1, which were not observed at JKT, and also provided a zero
point for NGC6649 and M25, which were only observed in non-photometric
conditions at the JKT.

\subsubsection{UKIRT} \label{sec:ukirtcal}

The standards stars observed at UKIRT were taken from the faint standards
list available from the UKIRT Web page. The standard stars were observed as
a mosaic of five frames, a central frame with an overlapping frame in each
direction. A 15$^{\prime\prime}$ aperture was used to determine the
magnitude of the standard. Three of the nights were photometric, night1,2
and 4 (16,17,19/6/97) so all the clusters were observed in photometric
conditions.

\begin{figure} 
{\epsfxsize=6truecm \epsfysize=6truecm \epsfbox[60 170 550
620]{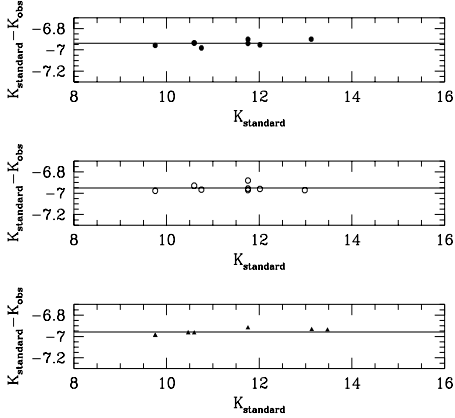}} 
\caption{The {\it K}-band zero
points for the three photometric nights at UKIRT. The top panel is for night
1, (16/8/97, filled circles), centre for night 2 (17/8/97, open circles) and the lower panel for night 4 (19/8/97, filled triangles).} 
\label{fig:kzero} 
\end{figure}

\begin{figure} 
{\epsfxsize=6truecm \epsfysize=6truecm \epsfbox[60 170 550
620]{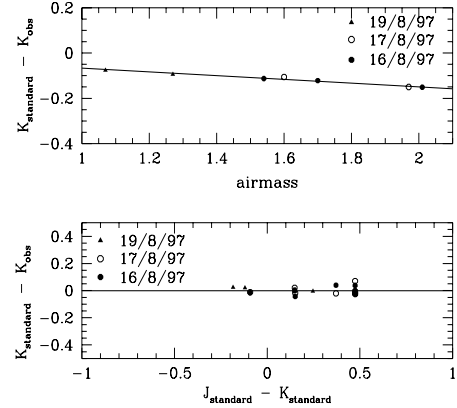}} 
\caption{The airmass
correction and colour equation for the {\it K}-band data taken at UKIRT. The symbols are the same as in Figure \ref{fig:kzero}.}
\label{fig:kaircol} 
\end{figure}

The zero points for each of the nights are shown in Figure \ref{fig:kzero}
and the airmass coefficient and colour equation are shown in Figure
\ref{fig:kaircol} and given below for each night.

\noindent16/06/1997; 

\begin{displaymath} K_{\rm{U}} = K_{\rm{std}} + 6.94 -
0.082sec(z) + 0.005(J_{\rm{std}} - K_{\rm{std}}) \pm0.024 \end{displaymath}
17/06/1997; 
\begin{displaymath} K_{\rm{U}} = K_{\rm{std}} + 6.95 - 0.082sec(z)
+ 0.005(J_{\rm{std}} - K_{\rm{std}}) \pm0.033 \end{displaymath} 
18/06/1997;
\begin{displaymath} K_{\rm{U}} = K_{\rm{std}} + 6.96 - 0.082sec(z) +
0.005(J_{\rm{std}} - K_{\rm{std}}) \pm0.037 \end{displaymath} 
where again the subscript std refers to the standard stars. The
difference between each of the calibrations is just a small change in the
zero point. The error on the airmass coefficient is $\pm$0.002 and the error on the colour term is $\pm$0.0003. The airmass coefficient agrees well with the values given in \citeasnoun{kris} and those found on the UKIRT web page. An aperture correction of -0.19$\pm$0.018 magnitudes is required for the magnitudes of the data frames measured using a 5$^{\prime\prime}$ aperture on each night. As the colour term is very small, it was assumed negligible in the {\it V-K:V} CMD's.

\subsubsection{Calar Alto and WHT} UKIRT faint standards and \citeasnoun{hunt}
standards were observed in order to calibrate the Calar Alto and WHT data.
The calibration was provided for us by Nigel Metcalfe as part of another project (see McCracken, Metcalfe and Shanks, in preparation, for more details). 

\begin{figure}
{\epsfxsize=8.5truecm \epsfysize=8.5truecm 
\epsfbox[60 170 550 620]{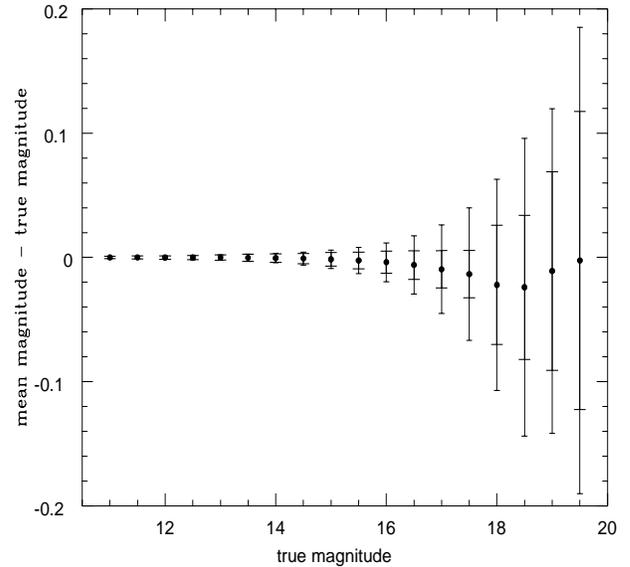}}
\caption
{An indication of the accuracy of the {\it U}-band photometry based on results for simulated stars (see text for details). The large errorbar indicates the total error (Poisson errors, crowding errors and also read noise), the smaller, wider errorbar indicates the Poisson error found as described in section \ref{sec:photometry}, paragraph 2. The total error at 17th magnitude is only $\pm$0.036 magnitudes and down to 18th magnitude the error is less than $\pm$0.1 magnitudes.}
\label{fig:phot}
\end{figure}

\subsection{Photometry}
\label{sec:photometry}

Automated aperture photometry was done using PHOT within IRAF's DAOPHOT
package. A small, 5$^{\prime\prime}$, aperture was used to minimise any
crowding problems. PHOT was also used to obtain the magnitudes of the
standard stars, the magnitude of each standard star was measured individually using a 15$^{\prime\prime}$  aperture.

First we establish the depth of the photometry. We define the limiting depth
of the observations to be where the Poisson error in the electron counts is
5\%, i.e. when $\sqrt{N_{obj} + N_{sky}}/N_{obj}$ = 0.05. The depth of the JKT data is 19.9 mags for a 1200s {\it U}-band exposure, 19.7 mags for a 360s {\it B}-band exposure and 20.4 mags for a 240s {\it V}-band exposure. The depths of the CTIO data are similar. The depth of a typical {\it K}-band exposure of 120s observed was around 19.2 magnitudes.

The accuracy of the zero points were tested by comparing our photometry to photoelectric observations in the literature. Table \ref{tab:resids} shows the residual when our magnitudes for typically 10
stars from the calibrated data frames (see section \ref{sec:cal}) are
compared to the magnitudes found from previous photoelectric studies. In all
cases except for NGC6067, the residual is small (less than 0.03 mags).
The errors in the comparison of our magnitudes to other photoelectric observations are slightly smaller than the errors in the zero point found from the standard stars in this work. This is probably because the stars considered when comparing the previous work were slightly brighter. 
Unfortunately for the cluster NGC6067 there is little photoelectric data
available and only 6 stars can be compared in the {\it B} and {\it V}-bands due to saturation and different parts of the cluster being observed. In the
{\it U}-band there are only 2 stars in common. Given that the offsets between
the work here and previous work are small for all the other clusters, we
assume that the zero point obtained for NGC6067 is accurate.

Finally, we test how the crowding of the field effects the photometry.
Twenty ``simulated stars'' of each magnitude shown in Figure \ref{fig:phot} were placed in a 300s {\it U}-band image of the cluster M25 then
PHOT was used to determine how well the magnitudes could be recovered. Shown in
Figure \ref{fig:phot} are the results. The x axis shows the true magnitude of the stars and the y
axis shows the deviation of the mean magnitude from the true magnitude. This plot shows the total error (Poisson errors, read noise and also errors due to crowding) at each magnitude (full errorbar) and the Poisson error, found as described above, as the smaller, wider, errorbar. The total error for this exposure of 300s is less than 0.03 magnitudes down to {\it U}=17 mags. For a {\it U}-band exposure of 1800s, assuming that the crowding errors remain the same and then
correcting the error shown in Figure \ref{fig:phot} for the reduced Poisson error, the total error is estimated to be $\pm$0.17 at {\it U}=20, reducing to $\pm$0.08 at 18th mag and $\pm$0.03 at 17th mag. Figure \ref{fig:phot} also shows that although the errors increase for fainter magnitudes there is no systematic trend.

\section{Reddening and Distance} 
\label{sec:rd}

\subsection{Method} 
\label{sec:method}

In order to work out the Cepheid Period-Luminosity (P-L) relation, the
distance to the cluster needs to be known. However, there is generally
significant dust absorption along the line of sight to the cluster which
must be corrected for. The method for determining both of these parameters
is done via ZAMS fitting to colour-magnitude and colour-colour diagrams. The
ZAMS used in this study is a combination of \citeasnoun{allen} for the
optical data, the intrinsic colours of near IR-band stars from the UKIRT Web
page \cite{tok} for the {\it K}-band data and nearby local stars taken from the
Strasbourg Catalogue to give an indication of the acceptable spread in the
ZAMS. The ZAMS of \citeasnoun{turnZAMS} and \citeasnoun{mermZAMS} have also
been tested and would give the same results as the ZAMS of
\citeasnoun{allen}.

The reddening is obtained using the {\it U-B:B-V} diagram which is 
independent of the distance to the cluster. The {\it U-B:B-V} diagram has always
been the favoured method for estimating the reddening; however, previously,
the accuracy of the reddening determination was limited by the depth of the
{\it U}-band photoelectric photometry. The improved U sensitivity of the current
generation of  CCD's should allow a potential improvement in the accuracy of
the reddening estimated from {\it U-B:B-V} diagrams and this is the route we
have adopted here.

The reddening law assumed in this work is from \citeasnoun{sharp}.

\begin{equation} \frac{{\rm E}(U-B)}{{\rm E}(B-V)} = 0.72 + 0.05{\rm E}(B-V)
\end{equation}

We choose to fit the ZAMS to the ridge-line of the  O and B stars 
rather than the least reddened envelope, because it helps take
account of differential reddening in some of the clusters. There is evidence
for differential reddening in NGC6823 and TR35 (see Figure \ref{fig:ubbv})
as the main sequence in the colour-colour diagram is substantially
broadened. By fitting to the centre of the data we measure the average
reddening for the cluster which we can then apply to colour-magnitude
diagrams which are uncorrected for differential absorption to obtain
distances. To determine the error we measure the standard deviation of
the O and B type stars from the ZAMS via least squares fitting. 
The errors quoted on the values for the reddening (see Table \ref{tab:rd}) are typically 0.1 mags and include any error in the calibration. In quite a
few cases, the ZAMS does not fit the colour-colour data well over the whole
range of B-V colours. This is particularly problematic in some cases and
these cases are discussed below. There is also the problem that the Cepheid
could have a different reddening to the cluster, caused by differential
reddening across the cluster or by the location of the Cepheid away from the
cluster. This is discussed further in section \ref{sec:PL}.

We use both the {\it B-V:V} and {\it V-K:V} colour-magnitude diagrams to determine
the cluster distance. {\it V-K:V} diagrams have the advantage that the slope of
the ZAMS is flatter than at {\it B-V:V}, possibly allowing more accurate distance estimates but {\it V-K:V} diagrams are available for only 7 of the clusters and the scatter in these diagrams is greater than in the {\it B-V:V} CMD's, particularly in the case of NGC6664 as there were difficulties aligning the {\it V} and {\it K} frames due to a shift in the telescope position half way through creating the {\it K}-band mosaic for this cluster. There are also few points on the {\it V-K:V} diagram for NGC7790 due to the small size of the IRCAM detector. This diagram was also made for us before the {\it V}-band observations for this cluster were available as a test for the feasibility of this project. The {\it V}-band data therefore comes from \citeasnoun{romeo}, however there is a good match between our {\it V}-band data and that of Romeo et al. (see Figure \ref{fig:7790comp}). The distance modulus which best fits the {\it B-V:V} CMD around the position of the A0V
stars, using the method of least squares, is taken to be the distance
modulus, $\mu_0$, of the cluster. The errors on the distance modulus were
found by measuring the standard deviation away from the ZAMS of A0V type
stars over the range -0.1$ \lsim ${\it B-V} $\lsim$0.1 in the dereddened ZAMS which covers a range of approximately 4 mags in {\it V}. The distance modulus found from the {\it B-V:V} CMD is then checked against the {\it V-K:V} CMD for the 7 clusters with such a CMD for consistency. In all cases the distance modulus found from the {\it B-V:V} diagram was consistent with the {\it V-K:V} CMD within the errors.

No attempt has been made to remove foreground and background stars. Only stars
which lie clearly off the main sequence (off in {\it B-V} by more than 1 mag for
example) were removed. There is no clear recipe for how to remove the
contaminating stars from the colour-magnitude and colour-colour diagrams so
the data shown in Figures \ref{fig:ubbv}, \ref{fig:bv} and \ref{fig:vk}
contain non-cluster members. The observations in this study only cover a field-of-view of $\sim 6^{\prime}$ and are pointed at the cluster centre so contamination may not be as much of an issue as for the wider field photographic plates. As an example, we discuss this further for the cluster NGC7790 in section \ref{sec:clus}.

All of the {\it U-B:B-V} diagrams, the {\it B-V:V} and the {\it V-K:V} diagrams are given
in Figures \ref{fig:ubbv}, \ref{fig:bv} and \ref{fig:vk} respectively and
our results for the reddening and distances obtained are given in Table
\ref{tab:rd} together with previous results as summarised by
\citeasnoun{LS2}. All of the clusters are individually discussed below.

\begin{table} \begin{tabular}{lcccc} Cluster & E({\it B-V})$_{\rm{clus}}$ &
E({\it B-V})$_{\rm{clus}}^{\rm{LS}}$ & $\mu_{\circ}$ & $\mu_{\circ}^{\rm{LS}}$\\
\hline 
NGC6649 & 1.37$\pm0.07$ & 1.35 & 11.22$\pm0.32$ & 11.28 \\ 
M25 &0.49$\pm0.08$ & 0.48 &  9.05$\pm0.43$ &  9.03 \\ 
NGC6664 & 0.66$\pm0.08$ &0.64 & 11.01$\pm0.37$ & 10.40 \\ 
WZ Sgr & 0.56$\pm0.20$  & 0.57&11.15$\pm0.49$ & 11.22 \\ 
Lynga 6 & 1.36$\pm0.17$ & 1.34 & 11.1$\pm0.45$&11.43 \\ 
NGC6067 & 0.42$\pm0.02$ & 0.35 & 11.17$\pm0.35$ & 11.13 \\
vdBergh1& 0.90$\pm0.18$ & 0.77 & 11.4$\pm0.65$ & 11.36 \\ 
Trumpler 35&1.19$\pm0.10$ & 0.92 & 11.3$\pm0.53$ & 11.56 \\ 
NGC6823 & 0.85$\pm0.09$&0.75 & 11.20$\pm0.55$ & 11.79 \\ 
NGC129 & 0.57$\pm0.06$ & 0.53*&10.90$\pm0.37$ & 11.24* \\ 
NGC7790 & 0.59$\pm0.05$ & 0.64* &12.70$\pm0.43$& 12.39* \\ \hline 
\end{tabular}

\caption{The values in the above Table come from the work here and from
\protect\citeasnoun{LS1}. The values marked * come from
\protect\citeasnoun{FW} as the clusters NGC129 and NGC7790 are not included
in the studies of Laney and Stobie.} \label{tab:rd} \end{table}

\subsection{Discussion of Individual Clusters} 
\label{sec:clus}

\textbf{NGC6649} has been studied previously by, for example, \citeasnoun{MVDB} and \citeasnoun{walk6649}. The agreement between the photometry of this study and the photoelectric data of \citeasnoun{MVDB} is good (see Table \ref{tab:resids}), with only small offsets in the {\it U} and {\it B}-band. \citeasnoun{MVDB} find E({\it B-V}) = 1.37 (no quoted error) for the reddening towards the cluster. The distance modulus $\mu_{\circ}$ from \citeasnoun{MVDB} is 11.15$\pm$0.7. 

\citeasnoun{turn6649} uses the photometry of \citeasnoun{MVDB} and that of \citeasnoun{talb} to study NGC6649 and find the cluster suffers from differential reddening. However for stars close to the cluster centre a value of E({\it B-V})=1.38 is appropriate. The distance modulus is found to be 11.06$\pm$0.03 when individual stars are dereddened.

\citeasnoun{walk6649} used {\it U, B} and {\it V}-band CCD data to study NGC6649. Agreement between the photoelectric data of \citeasnoun{MVDB} and \citeasnoun{walk6649} was found to be better than 0.03 mags in the {\it V}-band. \citeasnoun{walk6649} do not measure the reddening of the cluster due to the claims of differential reddening by Turner. \citeasnoun{walk6649} deredden each star individually to find a distance modulus of 11.00$\pm$0.15.

Rather than correct for differential reddening, we fit the ZAMS line to the centre of the {\it U-B:B-V} and {\it B-V:V} diagrams to try and measure the average values of the reddening. We find E({\it B-V})=1.37$\pm$0.07 and $\mu_{\circ}$=11.22$\pm$0.32, both values are consistent with previous work.
The distance modulus used in \citeasnoun{LS2} is slightly higher, 11.278 and the reddening,E({\it B-V}), slightly lower, 1.35 but again these values are well within the errors. 

\citeasnoun{barrell} found the radial velocity of the Cepheid V367 Sct to be -20$\pm$6 kms$^{-1}$ and the radial velocity of the cluster NGC6649 to be -14$\pm$5 kms$^{-1}$. This is taken to be evidence for the cluster membership of the Cepheid.

\textbf{M25} The photometry used in this study and that of \citeasnoun{sandm25} is compared in Figure \ref{fig:zero} and \ref{fig:col}. The agreement in all the wavebands is good, less that 0.03 mags different from that of Sandage.

M25 has a {\it U-B:B-V} diagram where the ZAMS fit is good over a wide range of {\it B-V} colours. 
The value for E({\it B-V}) of 0.49$\pm$0.08 is in excellent agreement with the work of \citeasnoun{sandm25} who obtained 0.49$\pm$0.05. \citeasnoun{johnm25} stated that E({\it B-V}) lay in the range 0.4 to 0.56 and \citeasnoun{SVDB} obtained E({\it B-V})=0.51$\pm$0.01. 

The distance modulus obtained for this cluster is 9.05$\pm0.43$  which is slightly higher than that of Sandage, who found $\mu_{\circ}$=8.78$\pm$0.15. The difference in the distance modulus is probably due to where to fit was made, we fit to the A type stars to obtain 9.05$\pm0.43$, Sandage's work contains brighter stars which may lie slightly off the main sequence. Wampler et al (1960) obtained 9.08$\pm0.2$ and \citeasnoun{SVDB} found $\mu_{\circ}$= 9.0$\pm$0.3 which agree well. There is also good agreement between our values and those used by \citeasnoun{LS2}, as given in Table \ref{tab:rd}.


\citeasnoun{feast} studied M25 by measuring radial velocities and spectra for stars within M25. He obtained radial velocities of around 4 kms$^{-1}$ for the
Cepheid U Sgr and for 35 stars in the cluster M25 indicating that U Sgr is a member of M25.


\textbf{NGC6664} has been studied previously by \citeasnoun{arpEV}. The agreement between his photometry and ours is good, with only small offsets between the two data sets as given in Table \ref{tab:resids}.

Figure \ref{fig:ubbv} shows that the cluster NGC6664 has an anomalous {\it U-B:B-V} diagram. This is a clear case where we fit the O and B type stars rather than trying to fit the F and G type stars as the F and G type stars maybe more affected by metallicity. The value for E({\it B-V}) is then 0.66$\pm$0.08 which is slightly larger than 0.6 obtained by \citeasnoun{arpEV}. No error is quoted by Arp. In the previous work by Arp, the observations were not deep enough to see if the anomalous shape of the {\it U-B:B-V} diagram would have been detected or not. Unfortunately the only previous source of photometry for more than a handful of stars is that of Arp so no other zero point comparisons can be made.

The distance modulus obtained is 11.01$\pm$0.37 which is larger than 10.8 from \citeasnoun{arpEV}, mostly due to the increased value for the reddening estimated here.

There is an larger discrepancy between the value of the distance modulus used by \citeasnoun{LS2} who quote a reddening of 0.64 but a distance modulus of 10.405. It is not clear why their distance modulus is so low as \citeasnoun{FW} use 10.88.

The radial velocity work of \citeasnoun{Kraft} shows EV Sct is a member of NGC6664. 

\textbf{WZ Sgr} The first problem with WZ Sgr is that its membership of an open cluster is questionable. \citeasnoun{turnWZ} discusses the membership of WZ Sgr to the cluster C1814-190 in some detail and concludes that the strongest evidence for membership of WZ Sgr to an open cluster comes from the fact that often the Cepheid in a cluster is around 4 magnitudes more
luminous than the B-type stars on the main sequence. This essentially means that the age of the Cepheid is consistent with the age of the cluster. \citeasnoun{FW} assume WZ Sgr to be a cluster member.

The cluster C1814-190 is also only comparatively sparsely populated with only around 35 members brighter than B$\sim$16 \citeasnoun{turnWZ}. There is also patchiness in the dust obscuration which would cause differential reddening \cite{turnWZ}. These two factors could possibly explain the odd shape of the {\it U-B:B-V} colour-colour diagram particularly in the range 0.5 $\lsim$ {\it B-V} $\lsim$ 1. Contamination from foreground and background stars could also be the source of the unusual {\it U-B:B-V} diagram. By measuring the reddening of the O and B type stars though we get a value for E({\it B-V})=0.56$\pm$0.20 which is in agreement with \citeasnoun{turnWZ} and consistent with E({\it B-V})=0.57 used by \citeasnoun{LS2}. 

The {\it B-V:V} diagram is surprisingly tight, giving a distance modulus of 11.15$\pm$0.49 which is again in agreement with \citeasnoun{turnWZ} who obtained 11.16$\pm$0.1. The distance modulus used by \citeasnoun{LS2} is slightly larger, 11.219 but again well within the errors of our result.

\textbf{Lynga 6} The photometry of Lynga 6 has been checked against that of \citeasnoun{moff} (shown in Figure \ref{fig:zero} and \ref{fig:col}) and also against that of \citeasnoun{VDBH}. Offsets between the different sets of photometry are less than 0.03 mags in each waveband.

Like NGC6649, Lynga 6 is also very heavily reddened which makes it very difficult to observe stars over a large range of {\it B-V} and {\it U-B} colours.
E({\it B-V}) is estimated to be 1.36$\pm$0.17 which is consistent with previous measurements by \citeasnoun{VDBH} who obtained 1.34$\pm$0.01, by \citeasnoun{madly6} who obtained 1.37$\pm$0.03 and the value of 1.34 used by \citeasnoun{LS2}.

The distance modulus obtained here is around 11.10$\pm$0.45 which is consistent with 11.15$\pm$0.3 obtained by \citeasnoun{walkly6} but is discrepant with the value used by \citeasnoun{LS2}. They have a distance modulus of 11.429 which agrees reasonably well with the value of \citeasnoun{madly6} who obtained ($V-M_V$)=16.2$\pm$0.5 for the apparent distance modulus which, if a value of 3.2 is assumed for the extinction coefficient, roughly implies 11.8$\pm$0.5 for the distance modulus (see section \ref{sec:PL} for more details of the extinction coefficient used here). The difference between the value of Madore and the value for the distance found here is that Madore tended to fit the edge of the ZAMS.

The Cepheid TW Nor lies close to the centre of the cluster Lynga 6 and 
has a very similar value of the reddening. This is taken as evidence of the membership of the Cepheid to the cluster \cite{walkly6}.

\textbf{NGC6067} There is very little photoelectric data for this cluster. The {\it B} and {\it V}-bands have been compared to the data of \citeasnoun{thack} but there are only 7 stars in common and in the {\it U}-band there are only two stars in common. The comparison shows that the zero point used here is at least consistent with previous the previous work. As the photometry obtained at CTIO for other clusters such as Lynga 6 and vdBergh1 agrees well with previous results, we have to assume that the photometry for NGC6067 is also good. \citeasnoun{walk6067} finds good agreement between their {\it B} and {\it V}-band CCD data and the photoelectric data of \citeasnoun{thack}.

NGC6067 has the lowest value for the reddening, of the clusters in this study, with E({\it B-V}) estimated as 0.42$\pm$0.02. The {\it U-B:B-V} diagram presented here has a main sequence which agrees fairly well with the ZAMS, although there is a spread around {\it B-V}=0.8. The value of 0.42 is slightly higher than previous values, \citeasnoun{coulson} obtained 0.35$\pm$0.1 and \citeasnoun{thack} obtained 0.33 with no quoted error.

As the reddening has been estimated to be slightly higher than previously, the distance modulus is also increased to 11.17$\pm$0.35 as opposed to 11.05$\pm$0.1 from Walker. However \citeasnoun{thack} found the distance modulus of the cluster to be around 11.3. This estimate is higher than our estimate, despite a smaller measured reddening, as the ZAMS fit was made to the edge of the CMD.

\citeasnoun{LS2} use 0.35 for the reddening and 11.13 for the distance modulus to the cluster.

The membership of the Cepheids to this cluster is discussed in detail by \citeasnoun{eggen}. Eggen noted that  
the Cepheid V340 Nor is centrally located in the cluster and has the same reddening as the cluster so is assumed to be a member. The Cepheid QZ Nor lies at a distance of two cluster radii out from the cluster centre \cite{walk6067} but is still assumed to be a cluster member by \citeasnoun{FW}

\textbf{vdBergh1} Our photometry is tested against the photoelectric data of  of \citeasnoun{arpCV} in Table \ref{tab:resids}. There is no significant offset between the two data sets and no evidence of any scale dependent error. \citeasnoun{turnCV} has compared their photometry to that of \citeasnoun{arpCV} and find good agreement.

The cluster vdBergh1 was only given a short exposure of 300s in {\it U} and 90s in {\it B} and {\it V}. The {\it U-B:B-V} diagram is therefore not very well populated at faint {\it U} magnitudes. There is a fair amount of scatter in the {\it U-B:B-V} diagram for this cluster, the average value is E({\it B-V})=0.9$\pm$0.18. This is larger than the value of previous estimates. \citeasnoun{turnCV} obtained a minimum value of E({\it B-V})=0.66, this clearly fits the the edge of the B-type stars. \citeasnoun{arpCV} obtained 0.76 (no error) but the spread in the {\it U-B:B-V} diagram is such that the larger value would also have been acceptable. 0.77 was used for the cluster reddening by \citeasnoun{LS2}

The distance modulus, $\mu_{\circ}$ obtained here is 11.4$\pm$0.65. This is larger than previous results, 10.94 obtained by \citeasnoun{arpCV} and 11.08 obtained by \citeasnoun{turnCV}. This is mostly due to the measurement of increased reddening. The value used by \citeasnoun{LS2} was 11.356.

The membership of CV Mon to the cluster has been determined by \citeasnoun{turnCV} using radial velocity measurements, evolutionary arguments and by it's location in the cluster.

\textbf{Trumpler 35} There is very little photoelectric data available for this cluster. We compare our photometry against the photoelectric observations of \citeasnoun{hoag} and find reasonable agreement in the {\it B} and {\it V}-bands. The agreement in the {\it U}-band is less good but as there is only one source of comparison and the JKT photometry appears to agree well for other clusters we suggest that the {\it U}-band data photometry is accurate. 

The data in the {\it U-B:B-V} diagram for the cluster Trumpler 35 (TR35) shows quite a large spread. This is probably caused by differential reddening. Rather than trying to correct for this, a mean value of E({\it B-V})=1.19$\pm$0.10 is taken for the reddening. Shown on the TR35 panel in Figure \ref{fig:ubbv} are two lines. The dashed shows the value of E({\it B-V})=1.03 taken from \citeasnoun{turn35} and the solid line shows the value of 1.19 adopted here. Using this, the distance modulus to the cluster is then estimated as 11.3$\pm$0.53 by fitting up the centre of the data. This is different to the value of \citeasnoun{turn35} who obtained 11.6$\pm0.16$ as Turner fitted the edge of the {\it B-V:V} diagram rather than the centre. The values for the reddening used by \citeasnoun{LS2} (E({\it B-V})=0.92, $\mu_{\circ}$=11.56) are in better agreement with those of \citeasnoun{turn35} rather than those obtained here.

The membership of RU Sct to Trumpler 35 using the reddening and evolutionary status of the Cepheid is discussed and supported by \citeasnoun{turn35}. However RU Sct does lie 15$^{\prime}$ away from TR35 \cite{turn35}.

\textbf{NGC6823} was only observed in non-photometric conditions. We use the photoelectric data of \citeasnoun{guetter} for calibration purposes. \citeasnoun{guetter} has compared his photometry with that of Hiltner (1956) and Hoag (1961) and finds that there are only small offsets of less than 0.04mags in {\it U-B} between the different data sets. \citeasnoun{turn6823} finds good agreement with the photometry of Hiltner (1956) which agrees well with the photometry of \citeasnoun{guetter} used for calibration here.

NGC6823 suffers from differential reddening \cite{turn6823}, perhaps to an even greater extent than Trumpler 35. Again shown in the panel for the cluster NGC6823 in Figure \ref{fig:ubbv} are two lines. One is for E({\it B-V})=0.53 taken from \citeasnoun{FW} and the other is E({\it B-V})=0.85 which fits the centre of the B-type stars. The value of E({\it B-V})=0.85$\pm$0.09 is assumed here. 

The distance modulus with this reddening is then 11.20$\pm$0.55, lower than the value of 11.81 found by \citeasnoun{turn6823} who again fitted the edge rather than the centre of the {\it B-V:V} diagram. \citeasnoun{LS2} assume a reddening of E({\it B-V})=0.44 and $\mu_{\circ}$=11.787. These values again agree fairly well with those of \citeasnoun{turn6823} rather than the values found here.

The Cepheid SV Vul is only thought to be associated with the cluster NGC6823 \cite{FW} as it lies a few arcmins away from the cluster centre.

\textbf{NGC129} was only observed in non-photometric conditions so previous work had to be relied upon for the calibration, as discussed in detail in section \ref{sec:jktcal}. The reddening for this cluster is E({\it B-V})=0.57$\pm$0.06. This value is slightly larger than previous values, \citeasnoun{turn129} obtained 0.47 but fits to the least reddened edge of the O and B stars rather than to the centre of these stars. \citeasnoun{arp129} found 0.53 (with no error) for the reddening which is in agreement with the value found here.

The distance modulus obtained assuming a value of 0.57 for the reddening is $\mu_{\circ}$=10.90$\pm$0.37. This value is lower than the value of 11.11 obtained by \citeasnoun{turn129} who again fits the edge of the main sequence rather than the centre. \citeasnoun{arp129} find 11.0$\pm$0.15 for the distance modulus of the cluster, in agreement with the value found here. DL Cas is not included in the study by \citeasnoun{LS2} due to the Northerly latitude of the cluster NGC129. The values for the reddening and the distance modulus used by \citeasnoun{FW} agree better with the values of \citeasnoun{turn129} rather than those obtained here.

\citeasnoun{Kraft} found a measurement of -14$\pm$3 kms$^{-1}$ for the radial velocity of the cluster NGC129. He found that the Cepheid itself had a radial velocity of -11km$s^{-1}$. Given that the error on any individual measurement is estimated to be around 1.5kms$^{-1}$, DL Cas is assumed to be a member of NGC129.

\textbf{NGC7790} The {\it U-B:B-V} diagram for this cluster appears quite clean
and well defined.  However, Fig. \ref{fig:ubbv}
shows that NGC7790 has a {\it U-B:B-V} diagram where the data poorly fits
the ZAMS line. We show one ZAMS shifted to fit the OB stars
which implies E({\it B-V})= 0.59$\pm$0.04 and another shifted to 
fit the F stars
which would imply E({\it B-V})=0.43$\pm$0.04. Most previous estimates 
are closer to that for the OB stars. This poor fit of the ZAMS to the 
{\it U-B:B-V} data in the case of this cluster  is particularly significant 
since it contains three Cepheids (see Table 1).

The {\it U-B:B-V} diagram for NGC7790 has a history of controversy. The 
original {\it UBV} photoelectric photometry  of Sandage (1958) of 33 
11$<${\it V}$<$15 stars was
criticised by Pedreros et al (1984). A check of 16 stars with the KPNO CCD seemed to confirm that Sandage's {\it U-B} and {\it B-V} colours were too blue by +0.075 and +0.025 mag respectively, although few details were given of errors etc. However we find excellent agreement between the work here and the photometry of Sandage, see Table \ref{tab:resids}, Fig. \ref{fig:zero} and Fig \ref{fig:col}. A direct comparison of the photoelectric observations of Sandage to the photographic observations in Pedreros (see Table 1 in Pedreros et al. 1984) implies that the differences in the colours of Sandage are too blue by $\sim$ 0.04 mags for both the {\it U-B} and {\it B-V} colours and when we compare our CCD photometry to 22 of the brightest photographically observed stars from Pedreros, we find similar colour differences of around 0.04 mags in both the {\it U-B} and {\it B-V} colours. 

The {\it U-B:B-V} diagram in \citeasnoun{ped} seemed to give the same sort of ill fitting ZAMS, throughout the range 0.3$<${\it B-V}$<$1.2, as found in this work. The suggestion was that the problem might lie in Sandage's photometry which Pedreros et al had used for calibration. However, we have tested various zero points for the {\it U-B} and {\it B-V} colours. With our own zero points for the colours we find the ill fitting ZAMS and even if we apply the corrections suggested by Pedreros to the offsets found between our photometry and the photographic data of Pedreros we still find an ill fitting ZAMS. 
\begin{figure} 
\begin{tabular}{c} {\epsfxsize=7.truecm \epsfysize=7.truecm
\epsfbox[60 170 550 620]{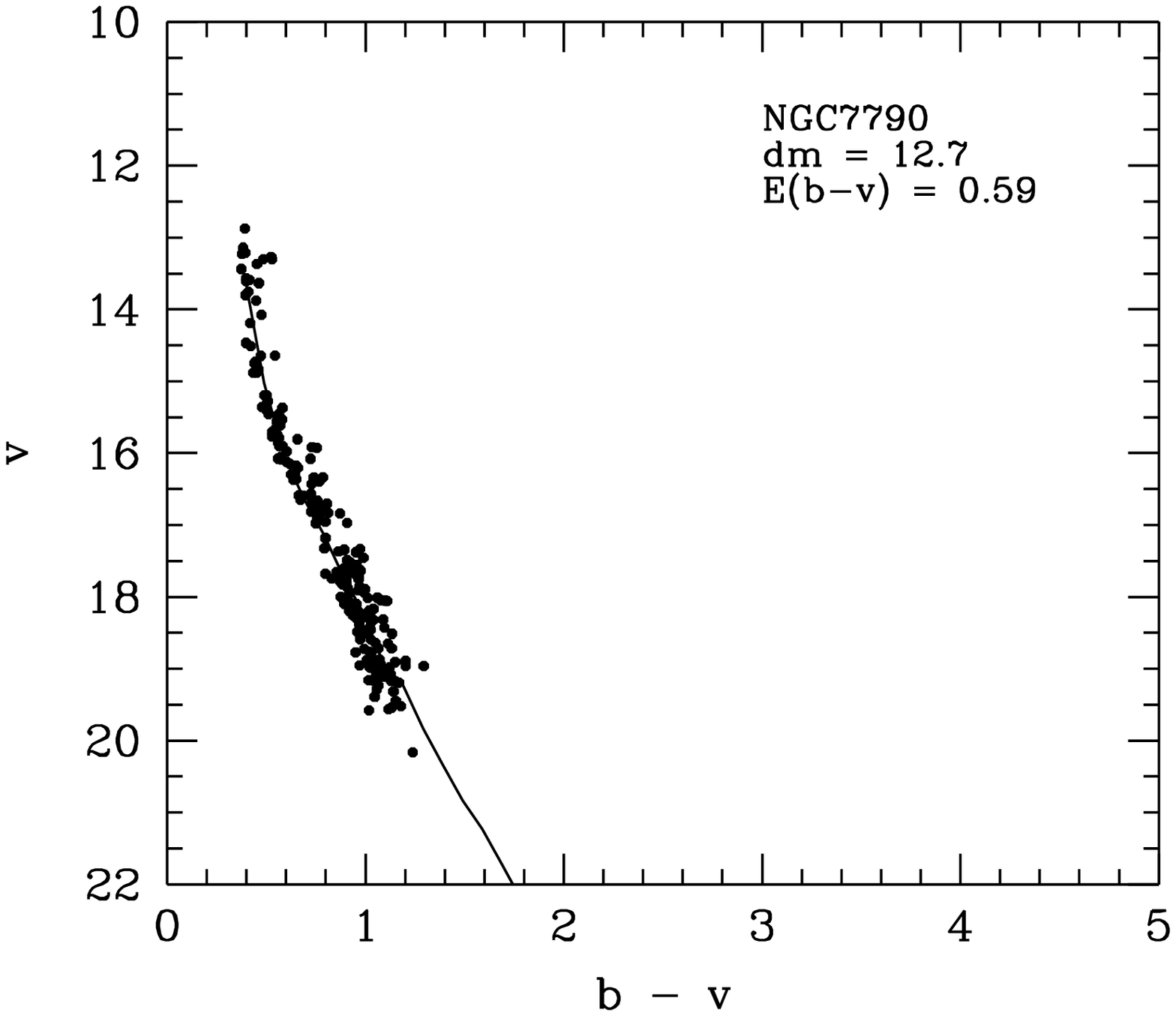}} \\
{\epsfxsize=7.truecm \epsfysize=7.truecm \epsfbox[60 170 550
620]{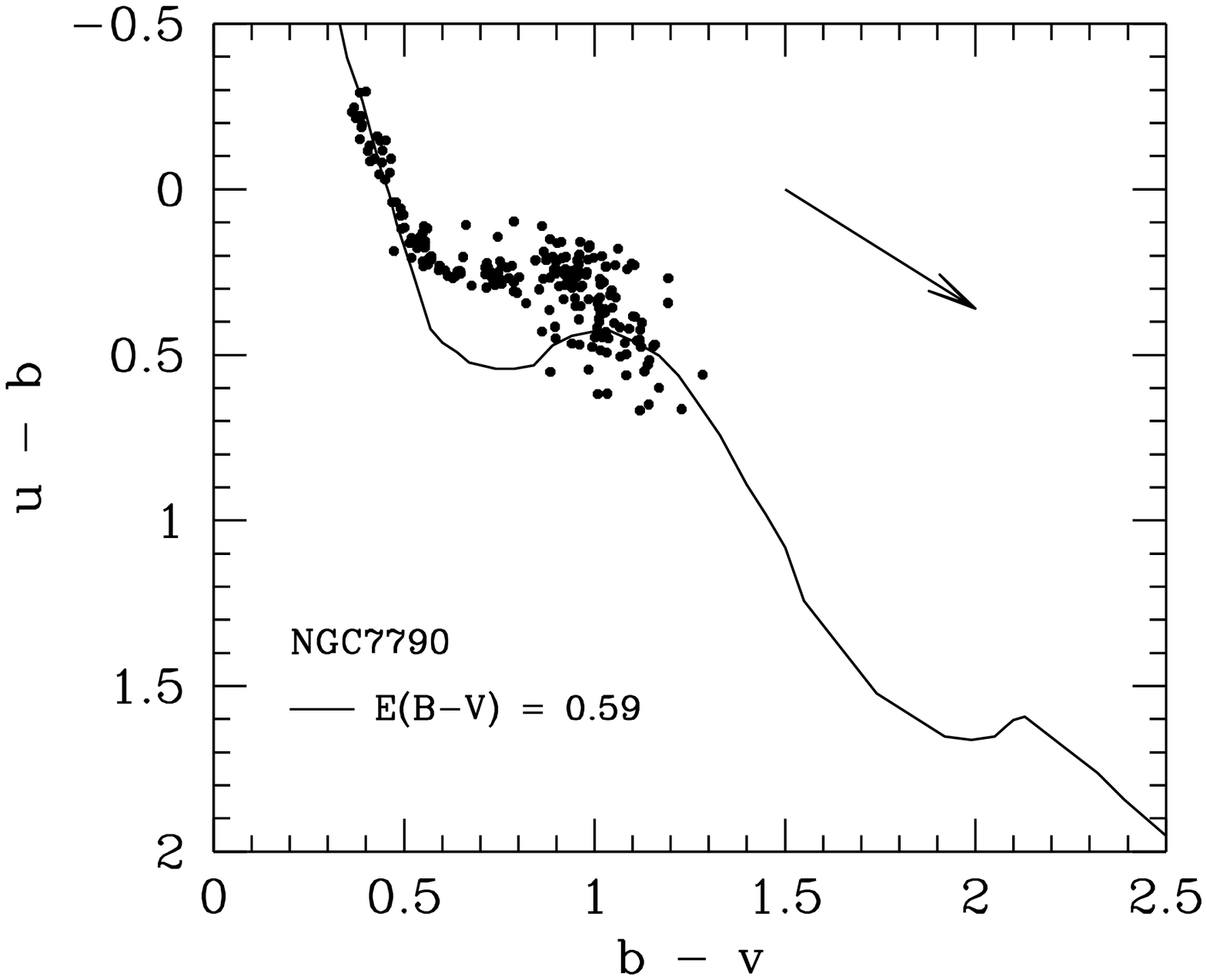}} \\ 
\end{tabular}
\caption{The top panel shows the {\it B-V:V} colour-magnitude diagram for the
cluster NGC7790 but with only the stars that fit the zero age main sequence
shown. The lower panel shows the same stars as they appear on the {\it U-B:B-V}
diagram. The U-V deficit is still clearly apparent among the stars that are
most likely to be members of the cluster. } 
\label{fig:memb} 
\end{figure}

\begin{figure}
{\epsfxsize=9truecm \epsfysize=8truecm
\epsfbox[60 170 660 620]{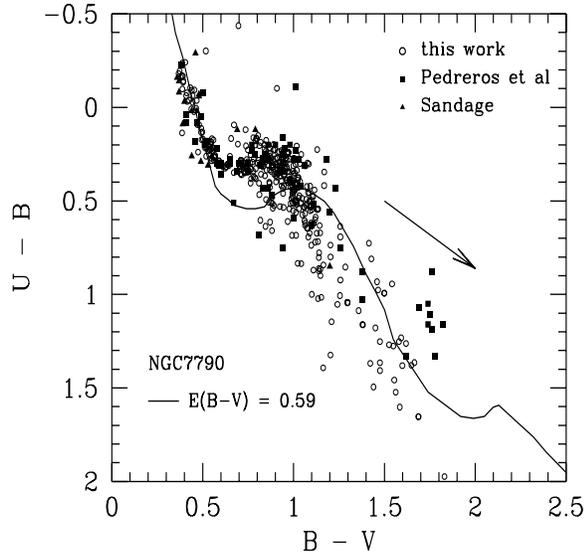}} 
\caption{The open circles show the photometry of this work. The solid triangles show the photometry of Sandage which does not go deep enough to test the ill fitting ZAMS. The results of \protect\citeasnoun{ped}(filled squares), though not as deep as those presented here, show the same effect as the photometry in this work and confirms the poor fit of the ZAMS {\it U-B:B-V} relation in NGC7790}
\label{fig:convince}
\end{figure}

Romeo et al (1984) used CCD data to obtain {\it BVRI} photometry for this cluster to {\it V}=20. In the absence of {\it U}, their only route to E({\it B-V}) was via fitting the shape of the {\it B-V:V} and {\it V-I:V} CMD and they obtained E({\it B-V})=0.54$\pm$0.04. They used the Sandage (1958) photometry for calibration in {\it B} and {\it V}, subject to the small re calibration in these bands by Pedreros et al. and checked against the {\it B,V}  photoelectric photometry of 10 stars by Christian et al (1985). They tested the faint photographic photometry of Pedreros et al and found scale errors at {\it B}$>$17 and {\it V}$>$15 in the sense that Pedreros et al were too bright. In {\it B-V}, however they claimed better agreement with {\it B-V}$_{Pedreros}$ being $\sim$0.1mag too red. A comparison of our photometry and that of Romeo suggests that there is a scale error for {\it B}$>$17 and {\it V}$>$16 but the extent of this is less than in the comparison of the Pedreros et al. data with the Romeo data, at {\it V}=17 our data is brighter than Romeo's by 0.1 mag whereas the Pedreros et al. data is brighter by 0.2 mags and similar differences are found in the {\it B} data. We conclude that the photometry in this study agrees very well with the photometry of \citeasnoun{sand7790} and is better agreement with the CCD data of \citeasnoun{romeo} than that of Pedreros, (see Fig. \ref{fig:7790comp} for more details). 

We have also attempted to check to see if contamination is causing the ill fitting ZAMS. Figure \ref{fig:memb} shows the {\it B-V:V} and the {\it U-B:B-V} diagrams for the cluster NGC7790. The {\it B-V:V}
diagram has been trimmed so that only the stars that lie very close to the
ZAMS remain. As these stars have the correct combination of distance and
reddening to lie almost on the main sequence then it is likely that they are
main sequence stars. The same stars are then used to produce the {\it U-B:B-V}
colour-colour diagram. The same {\it UV} deficit around the F-type stars that
appears in Figure \ref{fig:ubbv}, when all the stars in the field are used,
appears in Figure \ref{fig:memb}. Figure \ref{fig:convince} shows the colour-colour diagram found from this work with the points from Sandage and Pedreros included. The Sandage points do not go deep enough to test the shape of the data but the Pedreros et al. points do and the poor match of the data to the ZAMS is seen. Therefore, we believe the ill fitting ZAMS to the {\it U-B:B-V} diagram is not caused by contamination from foreground or background stars or by errors in the photometry but is a real feature in the data. 

Thus our estimate of the NGC7790 reddening based on the {\it U-B:B-V} colours of OB stars, E({\it B-V})=0.59$\pm$0.05, is between the E({\it B-V})=0.52 $\pm$0.04 of Sandage (1958) and the E({\it B-V})=0.63$\pm$0.05 of Pedreros et al(1984) who used similar techniques. 

Assuming the estimate of E({\it B-V})=0.59$\pm$0.05 from the OB stars, the
distance modulus which fits  {\it V:B-V}  is then $\mu_{\circ}$=12.72$\pm$0.11 which
is close to the original value of 12.8$\pm$0.15 found by Sandage(1958),
although this is in the opposite direction that would be expected from the
difference in reddening and slightly more than 12.65 obtained by
\citeasnoun{romeo} which is roughly in line with their obtaining
E({\it B-V})=0.54 for the   reddening. \citeasnoun{ped} obtained only 12.3 for
the distance modulus, no fits are shown in the paper and the reason they
obtain this low value is unclear. The values quoted by \citeasnoun{LS2}
are similar to those of \citeasnoun{ped}.

\citeasnoun{sand7790} states that the membership of CF Cas, CEa Cas and CEb
Cas to NGC7790 is almost certain due to the position of the Cepheids on the
CMD.

\begin{table*} \begin{tabular}{lccccccccccc} Cepheid & Log(P)
&E({\it B-V})$_{\rm{ceph}}$ & $<$V$>$ & $<$K$>$ & M$_{V}$ & M$_{K}$ &
M$_{V}^{\rm{LS}}$ & M$_{K}^{\rm{LS}}$  &  $\Delta(\rm{M}_{V})$  & $\Delta(\rm{M}_{K)}$ \\ \hline 
V367  & 0.799 & 1.27 & 11.604 & 6.662 &-3.789 & -4.937 & -3.755 & -4.992 & -0.034 &  0.055 \\ 
U Sgr & 0.829 & 0.45 & 6.685 & 3.952 & -3.841 & -5.232 & -3.781 & -5.214 & -0.060 & -0.018 \\ 
EV Sct & 0.490 & 0.62 & 10.131 & 7.028 & -2.877 & -4.164 & -2.191 & -3.553 & -0.686 & -0.611 \\ 
WZ Sgr & 1.339 & 0.50 & 8.023 & 4.565 &-4.810 &-6.738 & -4.657 & -6.789 & -0.153 &  0.051 \\ 
TW Nor & 1.033 & 1.24 & 11.670& 6.319 & -3.560 & -5.156 & -3.797 & -5.411 &  0.237 &  0.255 \\ 
QZ Nor & 0.730 & 0.27$\sharp$ & 8.866 & 6.662 & -3.164 & -4.634 & -3.128 & -0.036 & -0.046 & -0.035 \\ 
V340 Nor & 1.053  & 0.38 & 8.375 & 5.586 & -4.056 &-5.699 & -3.797 & -5.640 & -0.259 & -0.059   \\ 
CV Mon & 0.731 & 0.84 & 10.306 & 6.576 & -3.821 & -5.072 & -3.305 & -4.896 & -0.516 & -0.176 \\ 
RU Sct & 1.294 &0.93$\sharp$ & 9.465 & 5.071 & -4.901 & -6.508 & -5.186 &-6.775 & 0.284 & 0.267 \\ 
SV Vul & 1.654 & 0.44$\sharp$ & 7.243 & 3.920 & -5.440 & -7.415 & -6.028 & -8.004 & 0.588 &  0.589 \\ 
DL Cas & 0.903* & 0.60 & 8.97* & 5.93$\dagger$ & -3.648 & -5.136 & -3.860* & -5.36$\dagger$ & -0.171 & 0.181  \\ 
CF Cas  & 0.688* & 0.55 & 11.14* & 8.01$\dagger$ & -3.341& -4.872 & -3.170* & -4.85$\dagger$ & -0.171 & -0.022  \\ 
CEa Cas & 0.711* &0.55 & 10.92* & NA & -3.561 &   NA & -3.390* & NA & -0.171 & NA \\ 
CEb Cas & 0.651*& 0.55 & 10.99* & NA & -3.493 &   NA & -3.330* & NA & -0.163 & NA  \\ 
\hline
\end{tabular}
 
\caption{Comparison between the work here and that of
\protect\citeasnoun{LS2}. * indicates where values are taken from
\protect\citeasnoun{FW} and $\dagger$ where the values are inferred from
\protect\citeasnoun{welch}. $\sharp$ indicates where the Cepheid reddenings
have been corrected (see text)} \label{tab:diffs} \end{table*}

\section{P-L Relation} \label{sec:PL}

Using the values for the distance modulus and the reddening towards the
cluster, we proceed to determine the P-L relation. As well as the reddening
and the distance modulus of the cluster, the apparent magnitude of each of
the Cepheids is required. Where possible these come from LS (1993, 1994) who
have high quality {\it V} and {\it K}-band measurements for most of the Cepheids in
this study. The clusters NGC7790 and NGC129 lie at northerly latitudes so
are unobservable from SAAO and so there are no magnitudes from Laney and
Stobie for these Cepheids. The {\it K}-band data for DL Cas and CF Cas comes
therefore from \citeasnoun{welch} and the {\it V}-band data and the periods are
taken from \citeasnoun{FW}.

The reddening obtained from the ZAMS fitting is that of the cluster OB
stars. \citeasnoun{SK} found that when the effect  of the colour difference
between the OB stars and the Cepheid is taken into account,

\begin{figure} 
{\epsfxsize=8.5truecm \epsfysize=8.5truecm \epsfbox[60 170
550 620]{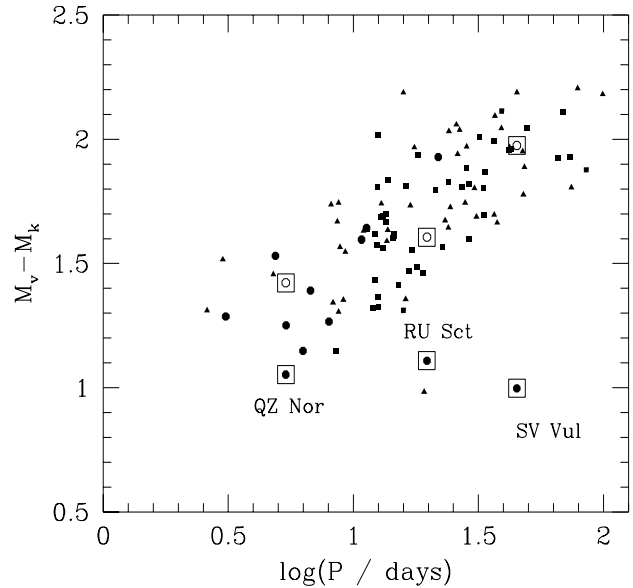}} 
\caption {The Galactic Cepheids are
shown by the solid circles, the LMC Cepheids by the triangles and the SMC
Cepheids by the squares, taken from \protect\citeasnoun{LS2}. The three solid 
circles in a box are QZ Nor, RU Sct
and SV Vul. The reddenings of these Cepheids are corrected to the values
given in \protect\citeasnoun{LS1}, indicated here by the boxed open
circles.} 
\label{fig:MvMk} 
\end{figure}

\begin{eqnarray} 
{\rm E}(B-V)_{\rm{ceph}}& = & {\rm E}(B-V)_{{\rm
clus}}[0.98- \nonumber \\ 
& & 0.09(<B_o>-<V_o>)_{{\rm ceph}}]
\label{eq:cephred} 
\end{eqnarray} gives a good approximation to the
reddening of the Cepheid. The values for $<B_o>-<V_o>)_{\rm{ceph}}$ 
come from \citeasnoun{FW}.

Figure \ref{fig:MvMk} shows the M$_{V}$-M$_{K}$ - Log(P) relation
with the reddenings found for the Galactic Cepheids (filled circles) found 
from the cluster reddenings obtained in this work via equation \ref{eq:cephred}.
The triangles show the same relation for the LMC Cepheids and the squares
are for the SMC Cepheids. These are taken from Tables 2 and 3 in \citeasnoun{LS2}.
Three of the Cepheids, (from left to right in Figure \ref{fig:MvMk})
QZ Nor, RU Sct and SV Vul seem to have the wrong M$_{\rm v}$-M$_{\rm k}$
colours for their periods. The clusters containing RU Sct and SV Vul 
suffer both from the presence of differential reddening and from the fact 
that the Cepheids lie at some distance away from the cluster. Figure \ref{fig:MvMk}
indicates that the reddening local to RU Sct and SV Vul may be somewhat 
different from the average value of the cluster reddening. We correct for 
this using the space reddenings given in \citeasnoun{LS1} which are more 
local to the Cepheids (see for example Turner 1980). Note that because 
there is difficulty obtaining the Cepheids true reddening from the cluster 
reddening for these two Cepheids we do not include them in the best sample 
(later in this section).
\citeasnoun{walk6067} notes that the Cepheid QZ Nor also lies
away from the centre of the cluster NGC6067, at a distance of two cluster
radii so again the reddening of the cluster may not be appropriate for
the reddening of the Cepheid. \citeasnoun{LS1} take the value for the
Cepheid reddening from \citeasnoun{coulson} of E({\it B-V}) = 0.265, derived
from {\it BVI}$_c$ reddenings. This is the reddening used to calculate the
position of the open circle in Figure \ref{fig:MvMk} and which we assume for
QZ Nor henceforth. However, the effect of changing the reddening of the cluster
to E({\it B-V}) = 0.265 changes the distance modulus to the LMC by less than
0.02, less than the error quoted here.

To determine the absolute magnitude of the Cepheid, the apparent magnitude
has to be corrected for reddening and distance. First of all the extinction
coefficient is required. We follow \citeasnoun{LS2} and use

\begin{equation} \Re({\rm ceph})= 3.07 + 0.28(B-V)_{\circ} + 0.04{\rm
E}(B-V)_{{\rm ceph}} \end{equation} to take into account the effect of
Cepheid colour on the ratio of total to selective extinction. The
Cepheid reddening comes from Eq. \ref{eq:cephred} except for the three
corrected values. Then the reddening free magnitudes of the Cepheids are

\begin{eqnarray} 
V_{\circ} & = &  V - \Re({\rm ceph}){\rm E}(B-V)_{{\rm
ceph}} \nonumber 
\\ 
K_{\circ} & = &  V_{\circ} - V + K + \frac{\Re({\rm ceph}){\rm E}(B-V)_{{\rm ceph}}}{1.1} 
\end{eqnarray} The expression for
$K_{\circ}$ has the form given above as the extinction coefficient in
the {\it K}-band is one tenth of that in the {\it V}-band. To  obtain
finally the absolute magnitude, the distance modulus, $\mu_{\circ}$, given
in Table \ref{tab:rd} has to be subtracted off.

The P-L relation can now be determined. We consider two samples, one where
we consider all the Cepheids available to us and another where the Cepheids
RU Sct and SV Vul are removed due to the problem of differential reddening
and the question of cluster membership. The zero points are obtained by
fixing the slope and obtaining the least squares solution using the Galactic Cepheids in this study. These are summarised in Table \ref{tab:zp}. The slopes that are considered are the slopes from
\citeasnoun{LS2} which are the best fitting slopes to all the Cepheid data
(Galactic open cluster Cepheids, LMC and SMC Cepheids) in their study. The
slopes are -2.874 in the {\it V}-band and -3.443 in the {\it K}-band. Also
considered is -2.81 in the {\it V}-band as this is the slope of the LMC Cepheids
and the slope used by \citeasnoun{FC}.

\begin{figure} 
\begin{tabular}{c} 
{\epsfxsize=7.7truecm \epsfysize=6.6truecm
\epsfbox[50 170 590 620]{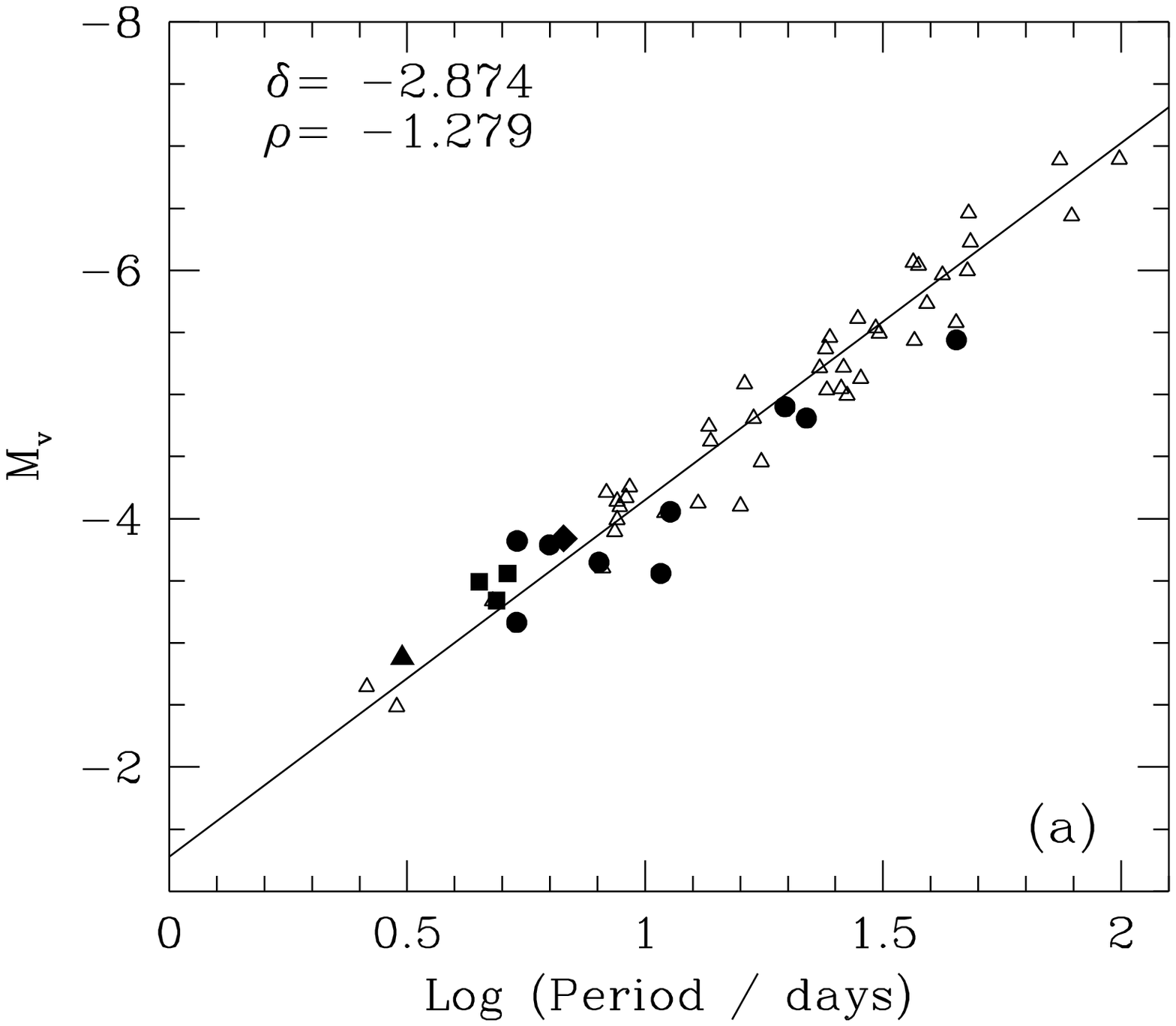}} \\
{\epsfxsize=7.7truecm \epsfysize=6.6truecm
\epsfbox[50 170 590 620]{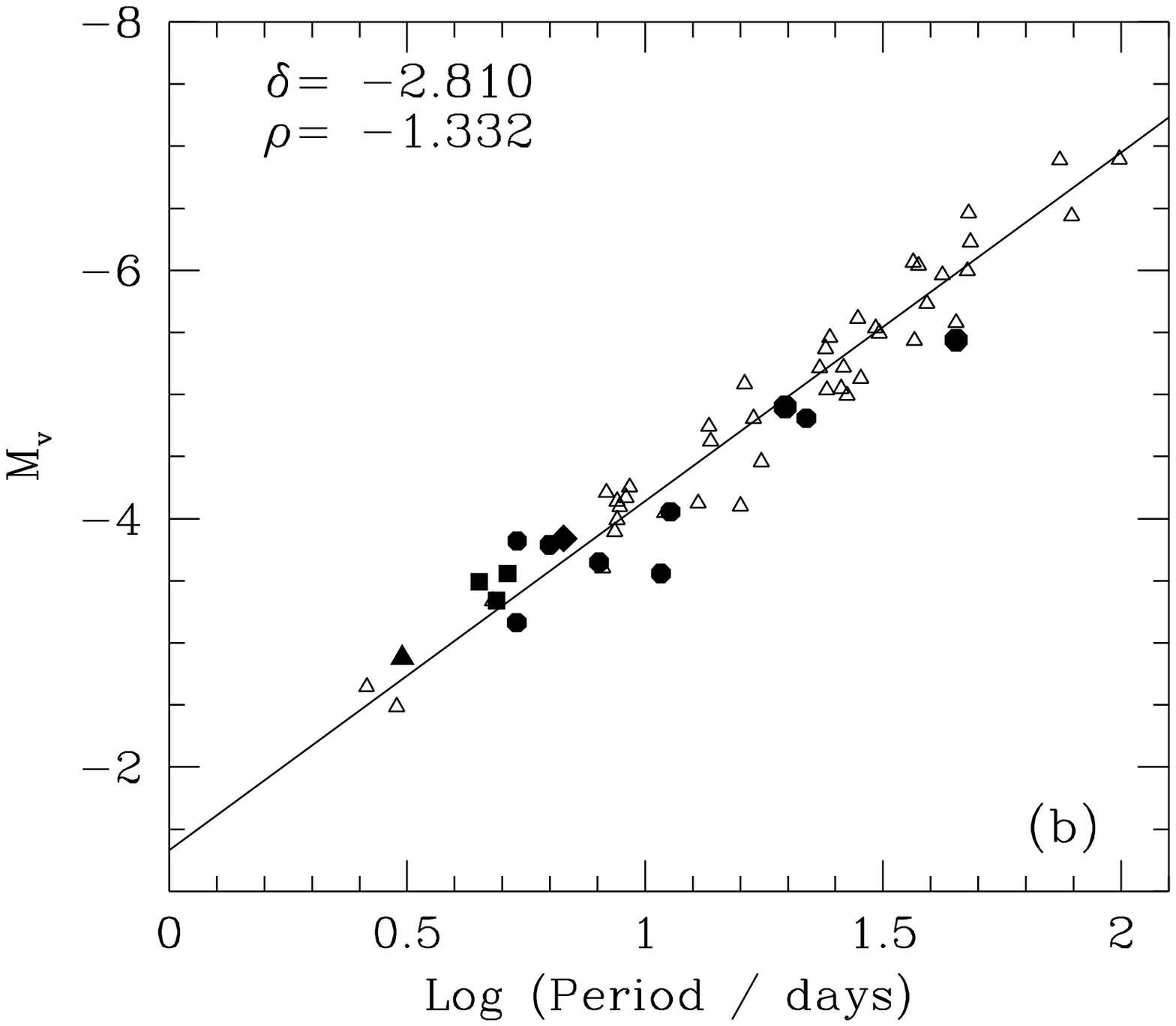}} \\ 
{\epsfxsize=7.7truecm \epsfysize=6.6truecm 
\epsfbox[50 170 590 620]{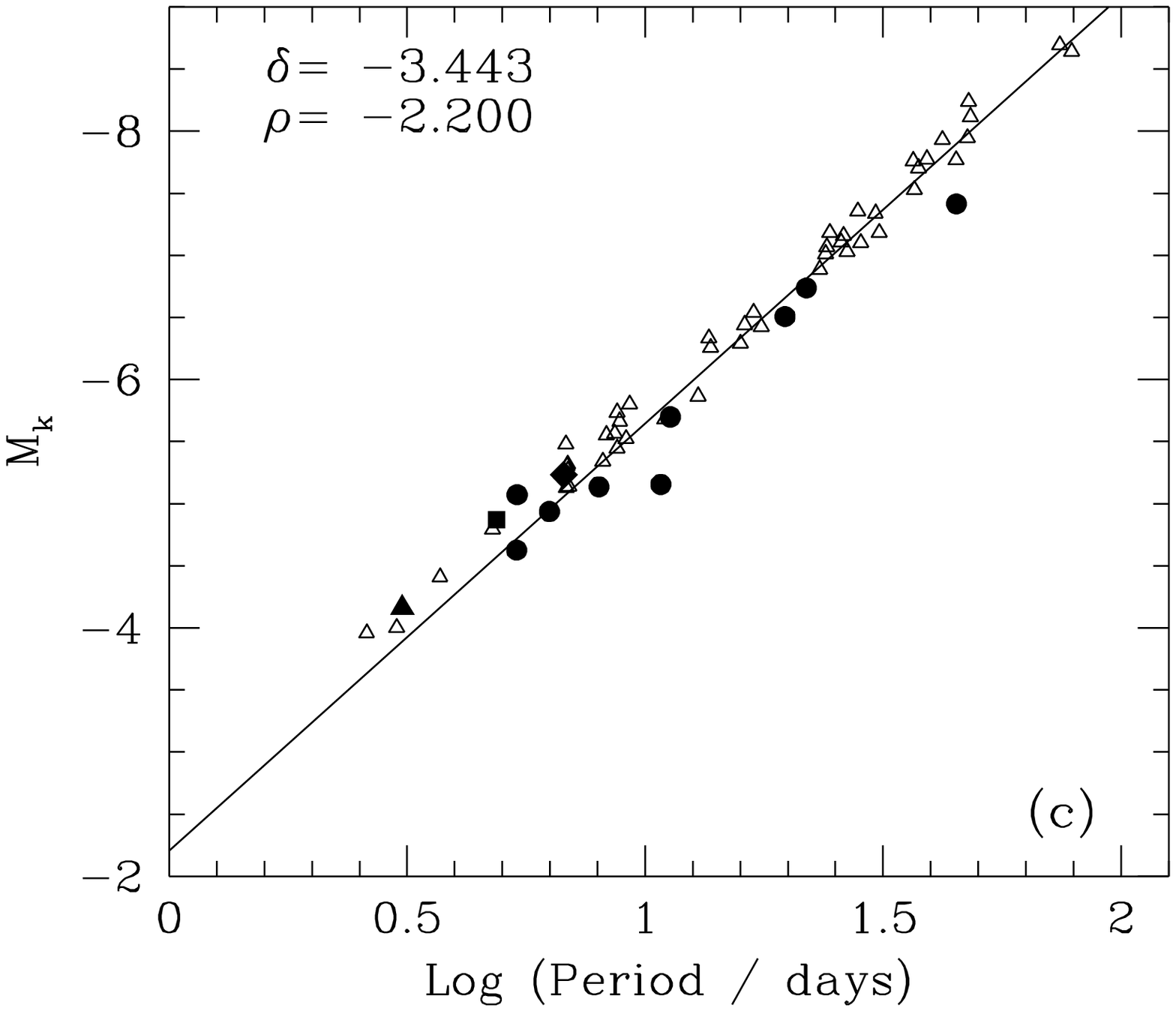}} \\
\end{tabular}
\caption{The Cepheid P-L relation. (a) shows the {\it V}-band P-L relation with a slope of -2.874 and the (b) shows the {\it V}-band P-L relation with a slope of -2.810. (c) shows the {\it K}-band P-L relation with a slope of -3.443. The solid symbols are the Galactic Cepheids, marked specifically are U Sgr (diamond), EV Sct (filled triangle), CF Cas, CEa Cas and CEb Cas (filled squares). There is no {\it K}-band data for the Cepheids CEa Cas and CEb Cas. The zeropoints and slopes are shown in each panel. The open triangles show the P-L relation for the LMC Cepheids. The periods and magnitudes are taken from Table 2 and 3 in \protect\citeasnoun{LS2} and the distance moduli to the LMC are the 'best' values from Table \ref{tab:zp} in this work.}
\label{fig:PL} 
\end{figure}

Once the slope and zero point of the PL relation is fixed, the distance
modulus to the LMC can be calculated. \citeasnoun{LS2} give the period and
the dereddened {\it V} and {\it K}-band magnitude, {\it V}$_{\circ}$ and {\it K}$_{\circ}$ of 45
LMC Cepheids. The distance to the LMC is then given by

\begin{equation} 
<\mu_{\circ}> = \frac{1}{\rm{N}} \sum_i^{\rm{N}} (m_{\circ} - \delta*\rm{log}(P)) - \rho 
\end{equation} where $m_{\circ}$ represents the
dereddened apparent magnitude in each waveband and $\delta$ and $\rho$ are the
values for the slope and zero point as given in Table \ref{tab:zp}. Our PL({\it V})
and PL({\it K}) relations are shown in Figure \ref{fig:PL} along with the equivalent
relation for the LMC Cepheids from \citeasnoun{LS2} using the distance moduli in
Table \ref{tab:zp} to determine the absolute magnitude.

\begin{table} \begin{tabular}{clccc} Band & Sample & Slope($\delta$) & Zeropoint($\rho$)
& $\mu_{\circ}$(LMC) \\ \hline 
{\it V}  & All(14)      & -2.874  &-1.229$\pm$0.34   &  18.53$\pm$0.036 \\ 
{\it V}  & All(14)      & -2.810  &-1.289$\pm$0.33   &  18.50$\pm$0.036 \\ 
{\it K}  & All(12)      & -3.443  &-2.148$\pm$0.30   &  18.42$\pm$0.019 \\ \hline 
{\it V}  & Best(12)     & -2.874  &-1.279$\pm$0.33   &  18.58$\pm$0.036 \\ 
{\it V}  & Best(12)     & -2.810  &-1.332$\pm$0.32   &  18.55$\pm$0.036 \\ 
{\it K}  & Best(10)     & -3.443  &-2.200$\pm$0.29   &  18.47$\pm$0.019 \\ \hline 
{\it V}  & logP$<$1.0(9)& -2.874  &-1.418$\pm$0.19   &  18.71$\pm$0.036 \\ 
{\it V}  & logP$<$1.0(9)& -2.810  &-1.465$\pm$0.19   &  18.68$\pm$0.036 \\ 
{\it K}  & logP$<$1.0(7)& -3.443  &-2.314$\pm$0.21   &  18.58$\pm$0.019 \\ \hline
\end{tabular} \caption{The zeropoints for the P-L relation and distance
modulus to the LMC} \label{tab:zp} \end{table}

Taking the value for the PL({\it V}) zeropoint for the best sample with the
-2.874 slope used by Laney \& Stobie (1994) gives $\rho$=-1.279$\pm$0.33
which is in good agreement with the value of $\rho$=-1.197$\pm$0.09 found by
these authors and implies a distance modulus of 18.57 for the LMC as
compared to 18.50 found by Laney \& Stobie. Figs. 8(a,b) show that the best
fitting zeropoint gives a somewhat poor fit to the majority of the Galactic
Cepheids which lie at log(P)$<$1.0. This is because the 12 Galactic Cepheids
in the best sample give a slope of $\delta$=-1.85$\pm$0.33 which is much
flatter than the LMC data which gives $\delta$=-2.79$\pm$0.1, close to -2.81.
Indeed, if only the 9 Galactic Cepheids with log(P)$<$1.0 are used, the
zeropoint rises to $\rho$=-1.418$\pm$0.19 and the LMC distance modulus would
rise to 18.70, indicating why the errors on the PL({\it V}) zeropoint are as large
as they appear in Table 5.

Our PL({\it V}) zeropoint is also  consistent with the zeropoint and LMC distance
obtained from an analysis of Hipparcos trigonometrical parallaxes of nearby
Galactic Cepheids by \citeasnoun{FC}. They obtained $\rho$=-1.43$\pm$0.1 for
the Galactic PL({\it V}) zeropoint for an assumed  slope of $\delta$=-2.81. This
can be compared to the  $\rho$=-1.332$\pm$0.32 obtained for our best sample
with the same slope. They used the same 45 Laney \& Stobie (1994) Cepheids
as used here to obtain a metallicity corrected LMC distance
modulus $\mu_{\circ}$=18.70$\pm$0.10. We note that  their semi-theoretical
metallicity correction to the LMC Cepheid {\it V} magnitudes  increases the
distance to the LMC, which is in the opposite sense to most empirically
determined estimates of the effects of metallicity on Cepheids (eg Kennicutt et al 1998). Subtracting their metallicity correction leads to an LMC distance modulus $\mu_{\circ}$=18.66$\pm$0.10 which can be
directly compared with our best value of $\mu_{\circ}$=18.55$\pm$0.036 from Table \ref{tab:zp}. We
conclude that our PL({\it V}) estimates the LMC distance modulus are between those
of Laney \& Stobie and Feast \& Catchpole but have too little statistical
power to discriminate between these previous estimates.

The {\it K}-band P-L relation is tighter for the LMC Cepheids and for the Galactic
Cepheids, see Figure \ref{fig:PL}(c), and the slopes are closer with the LMC Cepheids giving
$\delta$=-3.27$\pm$0.04 and the best sample of Galactic Cepheids  giving
$\delta$=-2.81$\pm$0.21. We note in passing that the slope of the Galactic
Cepheid PL({\it K}) relation is now much flatter than the $\delta$=-3.79$\pm$0.1
slope found in the Galactic Cepheid sample of Laney \& Stobie.  Assuming the
-3.443 slope used by Laney \& Stobie our best sample in Table 5 gives a
PL({\it K}) zeropoint of $\rho$=-2.200$\pm$0.29 which implies an LMC distance of
$\mu_{\circ}$=18.47$\pm$0.29 which remains in good agreement with the value
$\mu_{\circ}$=18.56$\pm$0.07 found by Laney \& Stobie. (1994). The smaller error
of Laney \& Stobie is due to their larger numbers of calibrators although it
must be said that many of their extra calibrators (8/12) are in associations
rather than clusters and frequently given half-weight in P-L fits. Indeed,
Hipparcos proper motion data has shown that one of their further cluster
Cepheids, S Nor, is also unlikely to be a member of its cluster  (Haguenau conference, September 1998).
Moreover, they have not included the 4 cluster Cepheids in NGC129
 and NGC7790. Therefore we believe that our result supercedes the Laney \& Stobie
result with our bigger error estimate perhaps being a more realistic
indication of the actual errors. Certainly in our best sample the biggest
changes in M$_K$ between Laney \& Stobie and ourselves, which contribute most
to  our 50\%  increased  scatter, are for EV Sct and TW Nor (see Table 4)
where the reasons for the distances used by Laney \& Stobie are unclear.

Finally,  as our overall estimate of the LMC distance, we take the average
of the PL({\it V}) and PL({\it K}) estimates in the best sample of Table 5 which gives
$\mu_{\circ}$=18.51$\pm$0.3. We conclude that although in the case of individual
clusters we have markedly improved the distance and reddening estimates, our
new estimates of the zeropoint of the PL relation and thus the distance to
the LMC are close to previous values.

\section{Discussion} 
\label{sec:interp}

\begin{figure} 
{\epsfxsize=8.5truecm \epsfysize=8.5truecm \epsfbox[125 210
560 600]{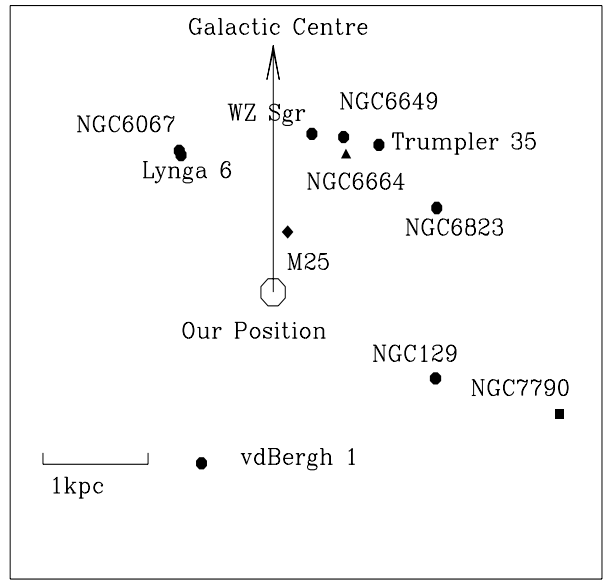}} 
\caption{The approximate positions of all the clusters considered here. The clusters NGC6664 (triangle), M25 (diamond), NGC129 and NGC7790 (square) are highlighted. These symbols are the same as in Figure \ref{fig:PL}} 
\label{fig:pos} 
\end{figure}

We now discuss the most intriguing new result in this study, which is that
the Solar metallicity ZAMS may not  always fit the {\it U-B:B-V} data in individual clusters.
This is not the first time an effect like this has been seen. \citeasnoun{turnros} saw
poorly fitting {\it U-B:B-V} ZAMS for the open cluster Roslund 3. Turner interpreted
this as evidence for the young, B-type stars having a cocoon of
circumstellar dust around them. This cocoon of dust then increases the
reddening of the B-type stars as compared to the F and G type stars, causing
the ill-fitting {\it U-B:B-V} ZAMS. The O and B type stars in the {\it U-B:B-V}
diagram of Roslund 3 show a large spread around the ZAMS as the amount of
excess dust would probably vary from star to star. Excess reddening could perhaps also be so strong that it was causing some O and B stars to be so reddened that they appeared as F type stars. However, the O and B type
stars in the {\it U-B:B-V} diagram for NGC7790 are very tight so the shape of the {\it U-B:B-V} diagram is unlikely to be caused by
excess dust. 


The next possibility we consider is that the effect might be due to 
stellar evolution, However, the CMD for the clusters look unevolved 
even at AOV as might be expected for clusters which have Cepheid variables
and are expected to be less than 10$^8$ years old.

We also considered whether the discrepancy between the main sequence fitted
distance and the Hipparcos parallax to the Pleiades could explain our result.
If it were assumed that all open clusters had roughly the same composition
then the different forms for {\it U-B:B-V} that we find might be taken as evidence
that the colours of main sequence stars may not be unique. This possibility has also been discussed as an explanation of the problem with the MS
fitted distance to the Pleiades \cite{floor} and if it proves relevant in that case it will certainly also be worthy of further consideration here.


If the reddening vector in {\it U-B:B-V} varied as a function of Galactic
position then this would also affect our results. However, at least in the 
case of NGC7790 it seems that for whatever relative shift in U-B and B-V,
the ZAMS still has the wrong shape to fit the observed colour-colour relation.

The final  possibility is that metallicity is affecting the F stars' U-B
colours in some of these clusters. Qualitatively there is some evidence
supporting this suggestion. First,  `line blanketing' is well known to
redden the U-B colours of metal rich stars at F and G and low metallicity
sub-dwarfs are known to show UV- excess as the reverse of this case (e.g.
\citeasnoun{camFG}).  Second, there is some suggestion that the cluster,
NGC7790, that shows a UV excess lies outside the solar
radius while NGC6664 which is  redder in U-B at F tends to lie inside (see
Figure \ref{fig:pos}). M25 which lies closest to the Sun also fits the solar
metallicity {\it U-B:B-V} diagram as well as any of the clusters. Given that
metallicity in the Galaxy is known to decrease with Galactocentric radius,
this is suggestive of a metallicity explanation. Many of the other clusters'
{\it U-B:B-V} diagrams are either too noisy due to differential reddening  (Tr35,
NGC6823) or too obscured to reach the F stars (NGC6649, Lynga 6, vdBergh1) to
further test this hypothesis. However, NGC6067 forms a counter-example to
any simple gradient explanation, since it seems to have a normal UBV plot
and lies inside the solar position This would have to be accommodated by
allowing a substantial variation on top of any average metallicity gradient.

However, quantitatively the case for metallicity is less clear. The size of
the UV excess seen is much larger in the case of NGC7790 than
expected on the basis of previous metallicity estimates of these clusters,
or of any measurement of the amplitude of the Galactic metallicity gradient.
Using the Fe/H  vs $\Delta$ U-B relations of \citeasnoun{carney} or \citeasnoun{camgrad} it
would be concluded that NGC7790 showed $\Delta$ U-B $\sim$ 0.2mag
which corresponds to Fe/H$\sim$-1.5. Thus clusters which on the basis of
their Main Sequences and the presence of Cepheids, must be less than 10$^8$
yr old, would be implied to have near halo metallicity. Previously, \citeasnoun{pantos} find Fe/H$\sim$-0.3 for these 2 clusters. Also \citeasnoun{frycar} find Fe/H=-0.2$\pm$0.02 for NGC7790 while finding Fe/H=-0.37$\pm$0.03 for NGC6664 based on spectroscopy
of the Cepheids in these clusters themselves. Also according to the
Galactocentric metallicity gradient which is usually taken to lie in the
range -0.02-0.1dex kpc$^{-1}$ \cite{rana}, there should only be on the
average $\Delta$ Fe/H $\sim$ 0.3 in the range of metallicity covering these
clusters.

On the other hand, it should be noted that Panagia \& Tosi's Fe/H estimates
are based on more poorly measured estimates of UV excess than those
presented here and also that there is little agreement between the
metallicity estimates of Fry and Carney and those of Panagia \& Tosi.
Measuring the metallicity of the Cepheids themselves as attempted by Fry
and Carney is difficult since the effective temperature is a function of the
light curve phase and a small difference in estimated temperature can make
a large difference in metallicity. Also the Galactocentric metallicity
gradient at least as measured for open clusters depends on relatively poor
U-B photometry at the limit of previous data from \citeasnoun{janes},
\citeasnoun{camgrad} and \citeasnoun{pantos}. In any case, it is well accepted
that the dispersion in metallicity around the mean gradient is indeed high,
with the range -0.6$<$Fe/H$<$+0.3 at the Solar position. Further \citeasnoun{geisler} using Washington photometry to estimate metallicity also found an
example of a cluster, NGC 2112, only  $\sim$ 0.8kpc outside the solar radius
with Fe/H=-1.2, although this cluster is older than those discussed here.

However, it would also seem that the tightness of the P-L relations in Fig. \ref{fig:PL}
could form a final argument  against the idea that NGC7790 has
Fe/H$\sim$-1.5. If the
metallicity of the cluster NGC7790 was really Fe/H$\sim$-1.5 then at given B-V, MS
stars would be sub-dwarfs with $\sim$1mag fainter absolute V magnitudes than
normal solar metallicity main sequence stars (see Cameron, 1984, Figs. 5,6). Thus since we 
have used a normal Main Sequence to derive the distance to NGC7790,
is it not surprising that the Cepheids in these clusters lie so tight  on the
P-L relation when they should be a magnitude too bright if the low metallicity
hypothesis is correct. The only way that the low metallicity hypothesis for
NGC7790 could survive this argument is if it were postulated that
{\it the effect of metallicity on Main sequence star magnitude and Cepheid
magnitude were the same - then the effect of   our derived  distance modulus
being $\sim$1 magnitude too high would be cancelled out by the fact  that the Cepheid is
actually sub-luminous by 1 magnitude because of metallicity which would leave
the Cepheid tight on the P-L relation as observed.} This might not be too
contrived if  a low metallicity Cepheid prefers to oscillate about its
subdwarf,  rather than solar metallicity, zero-age luminosity (at fixed effective temperature) on the Main Sequence. 
This would lead to a strong implied metallicity effect on the Cepheid PL({\it V})
and PL({\it K}) zeropoints; the implication would be that $\frac{\delta M}{\delta Fe/H} \sim$0.66 in the sense that lower metallicity Cepheids are fainter. This
coefficient is within the range that has been discussed for the empirical
effects of metallicity on Cepheids by \citeasnoun{kenn} and \citeasnoun{gould} although the
most recent work by \citeasnoun{kenn} appears to give a lower value of
$\frac{\delta M}{\delta Fe/H} \sim$0.24$\pm$0.16 again in the same sense.

The immediate effect on the distance to the LMC with Fe/H=-0.3 is that our
estimate of its distance modulus would decrease from 18.5 to 18.3. However,
since all that is determined at  the LMC is the slope of the P-L relation,
then the zeropoints we have derived in Table 5  from the Galactic Cepheids
would still refer the P-L relation to the Galactic zeropoint. The ultimate
effect on H$_{\circ}$ would then be decided by the metallicity of the Cepheids, for
example, in the galaxies observed by the  HST for the Distance Scale Key
project \citeasnoun{ferr}, by \citeasnoun{tanvir} in the case of the Leo I Group and by
\citeasnoun{saha} in the case of SNIa.  \citeasnoun{zarit} have measured metallicities
in these galaxies already but if the dispersion in Cepheid metallicity is as
large as it is implied to be in the Galaxy then there may be some signature in
a wider dispersion in the Cepheid P-L relations in at least the high
metallicity cases. The possibility of detecting this signature is currently
being investigated (Shanks et al 2000 in prep.)


Obviously the most direct route to checking the metallicity explanation for
the anomalous behaviour seen in the {\it U-B:B-V} diagrams is to obtain
medium-high dispersion spectroscopy for a sample of F stars in 
NGC7790 and NGC6664 to determine the metallicity directly for these main
sequence stars. Currently proposals are in to use WHT ISIS spectrograph for this purpose.


\section{Conclusions} \label{sec:conc}

We have presented colour-colour diagrams and colour-magnitude diagrams of a
sample of galactic clusters which contain or are associated with Cepheids.
All the clusters have been observed using similar methods and the data
reduction and extraction has also been done with similar techniques. The use
of the improved {\it U}-band data has allowed powerful new checks of previous E({\it B-V}) estimates over a wide range of magnitudes.
In order to estimate the reddenings and
distance moduli, we have fitted all the clusters in the same manner and have not
attempted to correct for differential reddening but instead taken a simpler
approach and fitted the average reddening value of the cluster. In most cases the differences that we have found between values for the reddening and distance
modulus are small and where there are significant differences these can
mostly be explained by comparing whether the ZAMS fit was made to the centre or to the edge of the colour-colour and colour-magnitude diagrams.

The Cepheid P-L relations found from fitting the best sample are
M$_V$=-2.81$\times$log(P)-1.332 and M$_K$=-3.44$\times$log(P)-2.20 and a distance
modulus to the LMC of 18.54$\pm$0.32 in the {\it V}-band and 18.46$\pm$0.29 in the
{\it K}-band giving an overall distance modulus to the LMC of 18.50$\pm$0.3,
ignoring  any possible effect of metallicity. These results for both the PL
relations and the LMC distance are consistent with the previous results of
Laney \& Stobie (1994) although the improved distances and reddenings have
increased the errors over what was previously claimed. These increased errors
mean that our result for the PL({\it V}) relation are also consistent with the
result of Feast \& Catchpole (1997) from Hipparcos measurements of Cepheid
parallaxes, although this gives rise to an LMC  distance modulus of $\mu_{\circ}$=
18.66$\pm$0.1 as opposed to our $\mu_{\circ}$=18.51$\pm$0.3.

With the improved {\it U}-band data, we find that for at least two of the clusters, the
data in the {\it U-B:B-V}  two-colour diagram is not well fitted by the solar
metallicity  ZAMS. One possibility is that  significant metallicity variations
from cluster to cluster  may be affecting the {\it U-B} colours of  F- and G-type
stars. The problem is that  the metallicity variations this would require are
much larger than expected for young, open clusters. More work is therefore
required to  determine the metallicity of the individual main sequence  stars
in each of the clusters  NGC7790 and NGC6664. If metallicity is
proven to be  the cause of the anomalous {\it U-B:B-V} relations, then it would imply that the Cepheid P-L relation 
in both the visible and the near-infrared is strongly affected by metallicity.

\vspace{-0.6cm}

\section*{Acknowledgments} FH acknowledges the receipt of a PPARC
studentship. We thank Floor van Leeuwen for useful discussions and we thank
Henry McCracken and Nigel Metcalfe for help with the Calar Alto data
reduction and Patrick Morris for the WHT reduction. We thank the PATT
telescope committee for the time on the UKIRT, JKT and WHT telescopes.
We thank CTIO for supporting this project and 
the Calar Alto time allocation committee.

\begin{figure*} 
\begin{tabular}{c}
 {\epsfxsize=5truecm \epsfysize=5truecm
\epsfbox[60 170 550 620]{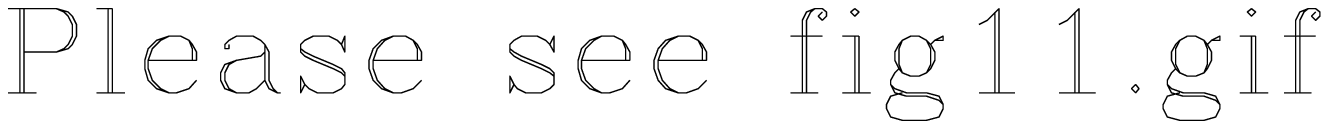}} 
\end{tabular}
\caption{The {\it U-B:B-V} diagrams for the clusters used in the study. The arrow
indicates the direction of the reddening vector.} \label{fig:ubbv}
\end{figure*}

\begin{figure*} 
\begin{tabular}{ccc} 
{\epsfxsize=5truecm \epsfysize=5truecm
\epsfbox[60 170 550 620]{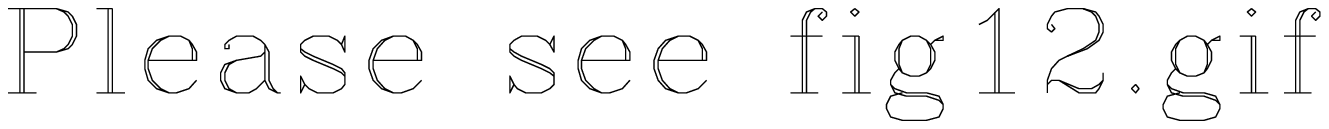}} 
\end{tabular}
\caption{The {\it V-B:V} diagrams for the clusters used in the study}
\label{fig:bv} \end{figure*}

\begin{figure*} \begin{tabular}{c} 
{\epsfxsize=5truecm \epsfysize=5truecm
\epsfbox[60 170 550 620]{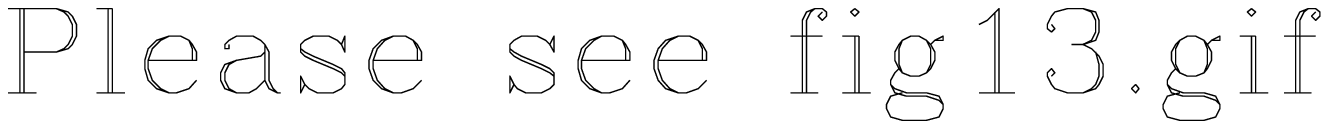}}
\end{tabular}
\caption{The {\it V-K:V} diagrams for the clusters used in the study}
\label{fig:vk} \end{figure*}

\begin{figure*} 
\begin{tabular}{c} 
{\epsfxsize=5truecm \epsfysize=5truecm
\epsfbox[60 170 660 620]{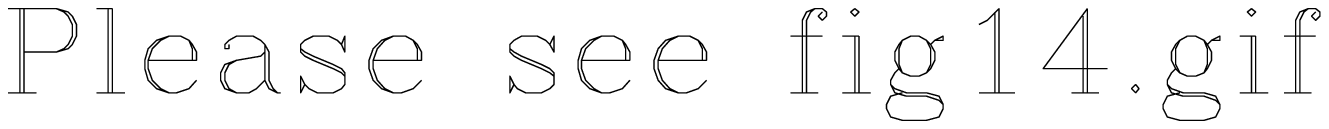}}
\end{tabular}
\caption{A comparison between the zero point compared in this work with the
zero point obtained in previous photoelectric studies. The full reference
for the comparison is given in Table \ref{tab:resids}. Only the clusters
where an independent zero points was obtained are shown.} \label{fig:zero}
\end{figure*}

\begin{figure*} 
\begin{tabular}{c} 
{\epsfxsize=5truecm \epsfysize=5truecm
\epsfbox[60 170 660 620]{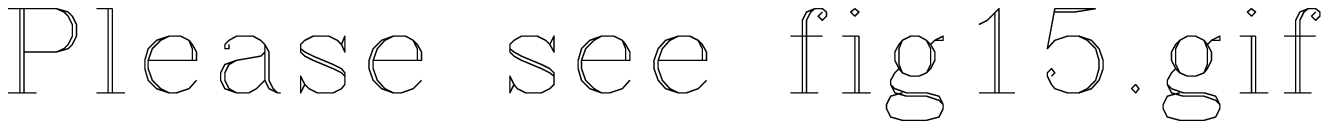}} 
\end{tabular}
\caption{A check for any colour dependent relationship between the work in
this study and previous photoelectric work. The full reference for the
comparison is given in Table \ref{tab:resids}. Only the clusters where an
independent zero points was obtained are shown.} \label{fig:col}
\end{figure*}

\begin{figure*}
\begin{tabular}{c}
{\epsfxsize=5truecm \epsfysize=5truecm
\epsfbox[60 170 660 620]{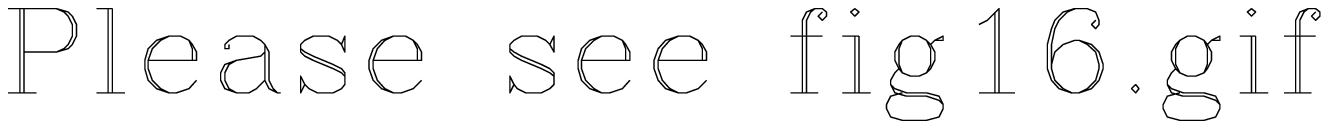}}
\end{tabular}
\caption{The zero points and check against colour for the clusters where previous work had to be relied upon for calibration purposes.}
\label{fig:other}
\end{figure*}

\begin{figure*}
\begin{tabular}{c}
{\epsfxsize=5truecm \epsfysize=5truecm
\epsfbox[60 170 660 620]{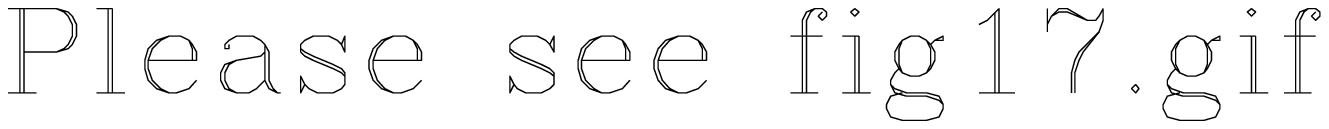}}
\end{tabular}
\caption{Comparison of all the different sources of photometry for NGC7790. Note there is no {\it U}-band data in the study by \protect\citeasnoun{romeo}}
\label{fig:7790comp}
\end{figure*}

\end{document}